\documentclass[fleqn,usenatbib]{mnras}

\usepackage{newtxtext,newtxmath}
\usepackage{subcaption}
\usepackage[flushleft]{threeparttable}

\usepackage[T1]{fontenc}

\usepackage{graphicx}	
\usepackage{amsmath}	
\usepackage{scalerel}
\usepackage{tikz}
\usetikzlibrary{svg.path}

\definecolor{orcidlogocol}{HTML}{A6CE39}
\tikzset{
  orcidlogo/.pic={
    \fill[orcidlogocol] svg{M256,128c0,70.7-57.3,128-128,128C57.3,256,0,198.7,0,128C0,57.3,57.3,0,128,0C198.7,0,256,57.3,256,128z};
    \fill[white] svg{M86.3,186.2H70.9V79.1h15.4v48.4V186.2z}
                 svg{M108.9,79.1h41.6c39.6,0,57,28.3,57,53.6c0,27.5-21.5,53.6-56.8,53.6h-41.8V79.1z M124.3,172.4h24.5c34.9,0,42.9-26.5,42.9-39.7c0-21.5-13.7-39.7-43.7-39.7h-23.7V172.4z}
                 svg{M88.7,56.8c0,5.5-4.5,10.1-10.1,10.1c-5.6,0-10.1-4.6-10.1-10.1c0-5.6,4.5-10.1,10.1-10.1C84.2,46.7,88.7,51.3,88.7,56.8z};
  }
}

\newcommand\orcidicon[1]{\href{https://orcid.org/#1}{\mbox{\scalerel*{
\begin{tikzpicture}[xscale=0.1,yscale=-0.1,transform shape]
\pic{orcidlogo};
\end{tikzpicture}
}{|}}}}

\usepackage{hyperref}   

\DeclareRobustCommand{\VAN}[3]{#2}
\let\VANthebibliography\thebibliography
\def\thebibliography{\DeclareRobustCommand{\VAN}[3]{##3}\VANthebibliography}

\usepackage{graphicx}	
\usepackage{amsmath}	

\newcommand{\tess}{{\it TESS}}

\newcommand{\gaia}{{\it Gaia}}
\newcommand{\JWST}{{\it JWST}}

\newcommand{\cheops}{{\it CHEOPS}}
\newcommand{\harpsn}{{\it HARPS-N}}
\newcommand{\sophie}{{\it SOPHIE}}
\newcommand{\cafe}{{\it CAFE}}

\newcommand{\kms}{km\,s$^{-1}$}
\newcommand{\ms}{m\,s$^{-1}$}

\newcommand{\rearth}{R$_{\oplus}$}
\newcommand{\mearth}{M$_{\oplus}$}

\newcommand{\vsini}{$v\sin{i}$}
\newcommand{\teff}{$T_{\rm eff}$}

\newcommand{\meh}{\mbox{$\rm [m/H]$}}

\newcommand{\logg}{$\log g$}

\newcommand{\logrhk}{$\mathrm{\log{R^{`}_{HK}}}$}

\newcommand{\refcom}{}
\newcommand{\Tstar}{HIP\,9618}
\newcommand{\TTstar}{TOI-1471}

\newcommand{\monotools}{\texttt{MonoTools}}

\newcommand{\TTeff}{$ 5609.0 \pm 33.0 $}

\newcommand{\TRs}{$ 0.9662 \pm 0.005 $}

\newcommand{\Tlogg}{$ 4.45^{+0.016}_{-0.033} $}

\newcommand{\Ttzerozero}{$ 1767.42089 \pm 0.00057 $}

\newcommand{\Ttzeroone}{$ 1779.1919 \pm 0.001 $}

\newcommand{\Tperiodzero}{$ 20.772907 \pm 2.3e-05 $}
\newcommand{\Tperiodzeroshort}{$ 20.77291 $}

\newcommand{\Tperiodone}{$ 52.563491 \pm 7.2e-05 $}
\newcommand{\Tperiodoneshort}{$ 52.56349 $}

\newcommand{\TMs}{$ 1.022^{+0.043}_{-0.076} $}

\newcommand{\Trorzero}{$ 0.03699 \pm 0.00037 $}

\newcommand{\Trorone}{$ 0.03171 \pm 0.00034 $}

\newcommand{\Trplzero}{$ 3.9 \pm 0.044 $}

\newcommand{\Trplone}{$ 3.343 \pm 0.039 $}

\newcommand{\TKzero}{$ 2.48 \pm 0.71 $}

\newcommand{\TKone}{$ 1.39 \pm 0.58 $}

\newcommand{\Tecczero}{$ 0.22^{+0.25}_{-0.15} $}

\newcommand{\Teccone}{$ 0.23^{+0.25}_{-0.16} $}

\newcommand{\Tomegazero}{$ 0.1 \pm 2.2 $}

\newcommand{\Tomegaone}{$ -0.1 \pm 2.1 $}

\newcommand{\TMpzero}{$ 10.0 \pm 3.1 $}

\newcommand{\TMpone}{$ <18.3 \pm 1.7 $}
\newcommand{\TMponeshort}{$ <18 $}

\newcommand{\Tsmazero}{$ 0.1438^{+0.0025}_{-0.0037} $}

\newcommand{\Tsmaone}{$ 0.2669^{+0.0046}_{-0.0069} $}

\newcommand{\TSinzero}{$ 54900^{+3100}_{-2100} $}

\newcommand{\TSinone}{$ 15930^{+890}_{-600} $}

\newcommand{\TTsurfpzero}{$ 663.4^{+9.1}_{-6.3} $}

\newcommand{\TTsurfpone}{$ 486.9^{+6.6}_{-4.6} $}

\newcommand{\Ttdurzero}{$ 0.2047^{+0.0015}_{-0.0014} $}

\newcommand{\Ttdurone}{$ 0.2731 \pm 0.0012 $}

\newcommand{\TTSMzero}{$ 166.0^{+74.0}_{-40.0} $}

\newcommand{\TTSMoneshort}{$ >42.0 $}

\newcommand{\TTIC}{306263608}

\newcommand{\TGAIA}{94468978202180352}
\newcommand{\Tra}{30.904272}
\newcommand{\Tdec}{21.280864}

\newcommand{\TTmag}{$ 8.5725 \pm 0.006 $}

\newcommand{\TGAIAmag}{$ 9.02543 \pm 0.00027 $}

\newcommand{\TKmag}{$ 7.559 \pm 0.021 $}

\title[Two Warm Neptunes transiting \Tstar{}]{Two Warm Neptunes transiting \Tstar{} revealed by TESS \& Cheops}

\author[H.~P.~Osborn et al.]{
\parbox{\textwidth}{
H.~P.~Osborn\textsuperscript{\hyperlink{affil_1}{1}, \hyperlink{affil_2}{2},\orcidicon{0000-0002-4047-4724}}\thanks{E-mail:hugh.osborn@unibe.ch} , 
G.~Nowak\textsuperscript{{\hyperlink{affil_3}{3}, \hyperlink{affil_4}{4}, \hyperlink{affil_5}{5}\orcidicon{0000-0002-7031-7754}}}, 
G.~Hébrard\textsuperscript{{\hyperlink{affil_6}{6}\orcidicon{0000-0001-5450-7067}}}, 
T.~Masseron\textsuperscript{{\hyperlink{affil_4}{4}, \hyperlink{affil_5}{5}\orcidicon{0000-0002-6939-0831}}}, 
J.~Lillo-Box\textsuperscript{{\hyperlink{affil_7}{7}\orcidicon{0000-0003-3742-1987}}}, 
E.~Pallé\textsuperscript{{\hyperlink{affil_4}{4}, \hyperlink{affil_5}{5}\orcidicon{0000-0003-0987-1593}}}, 
A.~Bekkelien\textsuperscript{{\hyperlink{affil_8}{8}}}, 
H.-G.~Florén\textsuperscript{{\hyperlink{affil_9}{9}}}, 
P.~Guterman\textsuperscript{{\hyperlink{affil_10}{10}}}, 
A.~E.~Simon\textsuperscript{{\hyperlink{affil_1}{1}\orcidicon{0000-0001-9773-2600}}}
% ...
}
\\
Affiliations are listed at the end of the paper in Appendix \ref{sec:affiliations}.
}

\date{Accepted XXX. Received YYY; in original form ZZZ}

\pubyear{2022}

\begin{document}
\label{firstpage}
\pagerange{\pageref{firstpage}--\pageref{lastpage}}
\maketitle

\begin{abstract}
\Tstar{} (HD\,12572, TOI-1471, TIC 306263608) is a bright ($G=9.0$\,mag) solar analogue. \tess{} photometry revealed the star to have two candidate planets with radii of \Trplzero{}\,$R_\oplus$ (\Tstar{}\,b) and \Trplone{}\,$R_\oplus$ (\Tstar{}\,c).
While the \Tperiodzeroshort{} day period of \Tstar{}\,b was measured unambiguously, \Tstar{}\,c showed only two transits separated by a 680-day gap in the time series, leaving many possibilities for the period.
To solve this issue, \cheops{} performed targeted photometry of period aliases to attempt to recover the true period of planet c, and successfully determined the true period to be \Tperiodoneshort{}\,d. 
High-resolution spectroscopy with \harpsn{}, \sophie{} and \cafe{} revealed a mass of \TMpzero{}\mearth{} for \Tstar{}\,b, which, according to our interior structure models, corresponds to a $6.8\pm1.4\%$ gas fraction.
\Tstar{}\,c appears to have a lower mass than \Tstar{}\,b, with \refcom{a 3-sigma upper limit of \TMponeshort{}\mearth{}.} 
Follow-up and archival RV measurements also reveal a clear long-term trend which, when combined with imaging and astrometric information, reveal a  low-mass companion ($0.08^{+0.12}_{-0.05} M_\odot$) orbiting at $26.0^{+19.0}_{-11.0}$ au.
This detection makes \Tstar{} one of only five bright ($K<8$\,mag) transiting multi-planet systems known to host a planet with $P>50$\,d, opening the door for the atmospheric characterisation of warm ($T_{\rm eq}<750$\,K) sub-Neptunes.
\end{abstract}




\section{Introduction}
The detection and characterisation of transiting exoplanets is currently the main driving force behind our rapidly expanding knowledge of exoplanets and exoplanetary systems. 
This is in part driven by the expanding capability to perform precise transmission spectroscopy, especially with JWST \citep[e.g.][]{TheJWSTTransitingExoplanetCommunityEarlyReleaseScienceTeam2022}.
The ability of the Transiting Exoplanet Survey Satellite \citep[\tess{};][]{Ricker2015} to find planets around bright stars has also greatly contributed, with \tess{} tripling the number of confirmed transiting planets around bright ($K<8$) stars \footnote{70/107 planets with $K<8$  \citep[\url{https://exoplanetarchive.ipac.caltech.edu}; accessed 2022-07-08; ][]{Akeson2013}}.

Since the majority of the sky is covered by \tess{} with only 27-day sectors, many long-period planets escape detection.
However, such planets are important. They are the ones least influenced by their parent stars, and therefore may maintain more primordial characteristics than their shorter-period siblings \citep{owen2019atmospheric}.
Their cooler atmospheres may permit the detection of different atmospheric molecules.
And their large Hill spheres mean these are the planets most likely to have stable moon or ring systems \citep[e.g.][]{dobos2021survival}.
Therefore, confirming longer-period transiting planets is key to expanding our knowledge of planetary formation and evolution and bridging the gap between Hot Jupiters and extrasolar systems more akin to the solar system.

While in much of the sky \tess{} is unable to catch consecutive transits of planets with orbital periods longer than 27\,d, it is able to observe the planetary transits of such planets -- either as single transits, or as non-consecutive "duotransits".
In the case of these duotransiting planets, which are being found in abundance during TESS' extended mission, we typically have a two year gap between transits.
Such cases are easier to solve than single transits as with two observed transits, the orbital period is limited to a discrete set of possible aliases. 
These can then either be searched using photometry or reduce the radial velocity phase space to specific periods.
This technique has proved extremely fruitful in the TESS extended mission, with detections of the periods of i.e. TOI-2257 \citep{schanche2022toi}, and TOIs 5152b \& 5153b \citep{ulmer2022two}.

While giant duotransiting planets can typically be redetected with ground-based photometry, for super-Earths and mini-Neptunes which produce only shallow transits, targeted space-based photometry is the more reliable way to redetect such a transit.
ESA's \cheops{} mission, a 30cm space telescope in low-Earth orbit and specifically designed for transit photometry, is the perfect instrument for this task.
\refcom{CHEOPS observations thus far been useful to reveal the true periods of warm mini-Neptunes, including TOI-2076\,c \& d \citep{Osborn2022}, TOI-5678\,b \citep{UlmerMoll2022}, HD\,15906\,c \citep{Tuson2022}, \& HD\,22946\,d \citep{Garai2022}.}

In this paper we report the discovery of a transiting multi-planet and multi-star system orbiting the bright (G=9.0, K=7.8) solar-like star \Tstar{}.
In section 2 we summarise the various data taken from \Tstar{}, including survey observations and targeted follow-up.
In section 3 we describe the analyses performed, including the derived stellar parameters and the final combined model used to derive planetary parameters.
Sections 4 \& 5 discuss the \Tstar{} system in context of the known exoplanetary systems and concludes.

\section{Data Description}

\subsection{TESS \& identification of two planetary signals}
\Tstar{} was initially observed by \tess{} in sector 17 (Oct 2019) at 2-minute cadence.
The data was processed by the The TESS science processing operations center \citep[SPOC;][]{jenkins2016tess}, which included aperture photometry, flagging poor-quality data, removal of trends associated with systematic and non-stellar sources \citep[presearch data conditioning;][]{Stumpe2012, Stumpe2014, 2012PASP..124.1000S}, and finally a search for transiting exoplanets \citep{2002ApJ...575..493J,2010SPIE.7740E..0DJ,2020TPSkdph}. SPOC processing revealed a threshold-crossing event – i.e. a candidate planet – which passed all Data Validation checks \citep{Twicken:DVdiagnostics2018,Li:DVmodelFit2019} and was subsequently manually vetted by a team of astronomers as TESS Object of Interest 1471.01 \citep{2021ApJS..254...39G}.
This initial candidate was the result of two transits seen in the S17 data, which were separated by 11.8\,d - the purported initial TOI period.

As with all TOIs, the target was passed to various groups in the \tess{} Follow-up Observing Program (TFOP) with the task of taking complementary observations to help confirm the candidate as a bona fide planet.
This included low-resolution spectroscopy to rule out large RV variations of an eclipsing binary as well as improving stellar parameters (see \ref{sect:TRES} \& \ref{sect:FIES}), high-resolution imaging to identify close-in blended stars which may be the source of the observed transits (see \ref{sect:Keck_palomar}, \ref{sect:Gemini}, \ref{sect:SOAR}, \& \ref{sect:SAI}), and ground-based photometry to confirm the purported ephemeris \& detect if its source is associated with the target star or a blended object (see \ref{sect:LCOGT}).

Intriguingly, follow-up photometry with LCOGT at 11.8\,d appeared to not reveal a transit at the TOI period as described in Section \ref{sect:LCOGT}.
Although this could have been a sign of Transit Timing Variations (TTVs) or even a blended source, this inspired a detailed inspection of the \tess{} lightcurve.
This analysis revealed that the two transits in fact had different depths \& durations, as well as finding a third transit in the S17 data which was present in the raw flux data but had been masked out during PDC detrending.
This extra transit was not consistent with the 11.8d TOI period, but was consistent in depth \& duration with the initial transit, suggesting a $\sim21$d period.
In this scenario the second transit seen in the original TESS light curve analysis was likely a monotransit from an outer planet orbiting \Tstar{}.

\tess{} subsequently observed \Tstar{} during the Ecliptic campaign in sectors 42 \& 43 (20th Aug -- 12th Oct 2021).
TESS full-frame image (FFI) data was initially reduced and processed by the Quick-Look Pipeline (QLP), which runs a complementary photometric extraction, detrending and transit search \citep{Huang2020}.
TESS data revealed a further 3 transits which QLP was able to use to detect the true period of TOI-1471.01 as \Tperiodzeroshort{}d.
After masking those transit events, QLP also detected a second transit of the outer planet candidate. This became TOI-1471.02. 
SPOC processing also revealed the true period of TOI-1471.01 when later processing S42 \& S43 data.

\subsubsection{Custom TESS detrending} 

In order to reveal the third transit in the S17 data, we re-extracted aperture photometry from the available TESS data and performed a custom detrending.
Starting from the \tess{} two-minute-cadence target pixel files, we extracted light curves from 20 different photometric apertures. 
For each light curve, we fit a model to the time series consisting of a linear combination of a basis spline (with breakpoints spaced every 1.5\,d to model long-timescale instrumental effects and stellar variability) and systematics parameters relating to the means and standard deviations of the spacecraft quaternion time series (and the squared time series) within each exposure \citep[see][]{Vanderburg2019}. Using a linear least squares technique (matrix inversion), we solved for the best-fit coefficients of our free parameters while iteratively excluding $3\sigma$ outliers from the fit until it reached convergence.
After calculating the best-fit systematics model for each aperture's light curve, we then subtracted it from the uncorrected light curve and identified the aperture that produced the light curve with the lowest photometric scatter. 
The best-performing aperture was roughly circular with a radius of about 3 pixels. For this aperture, we calculated the contribution of flux from stars other than the target and found that contamination was negligible (less than 0.5\%, much smaller than the uncertainty on the planet's transit depth) so we did not apply a correction for diluting flux.

\subsection{LCOGT}\label{sect:LCOGT}
Ground-based photometry was performed using the Las Cumbres Observatory Global Telescope \citep[LCOGT;][]{Brown:2013}.
The first observation was performed using the 1.0\,m node at McDonald Observatory on UTC 2020 January 13 to follow-up the ingress of the initially proposed orbital period of TOI-1471.01 at 11.767~d.
Two later observations were performed on the nights of UTC 2022 September 4 \& 2022 September 8 with the 1.0\,m telescopes at CTIO and SAAO respectively in order to confirm the periods found by \tess{} and \cheops{} observations.

All timeseries used the Pan-STARRS $z$-short band and used the {\tt TESS Transit Finder}, which is a customized version of the {\tt Tapir} software package \citep{Jensen:2013}, to schedule our transit observations. The 1\,m telescopes are equipped with $4096\times4096$ SINISTRO cameras having an image scale of $0\farcs389$ per pixel, resulting in a $26\arcmin\times26\arcmin$ field of view. The images were calibrated by the standard LCOGT {\tt BANZAI} pipeline \citep{McCully:2018}. The telescopes were moderately defocused to attempt to improve photometric precision resulting in typical TOI-1471 full-width at half-maximum of $3\farcs7$. Photometric data were extracted using {\tt AstroImageJ} \citep{Collins:2017} and circular photometric apertures with radii $7\farcs0$. The TOI-1471 photometric aperture included most of the flux of the delta TESS magnitude 7.382 Gaia eDR3 neighbor $3\farcs35$ southwest of TOI-1471. 

\begin{figure}
	\includegraphics[width=\columnwidth]{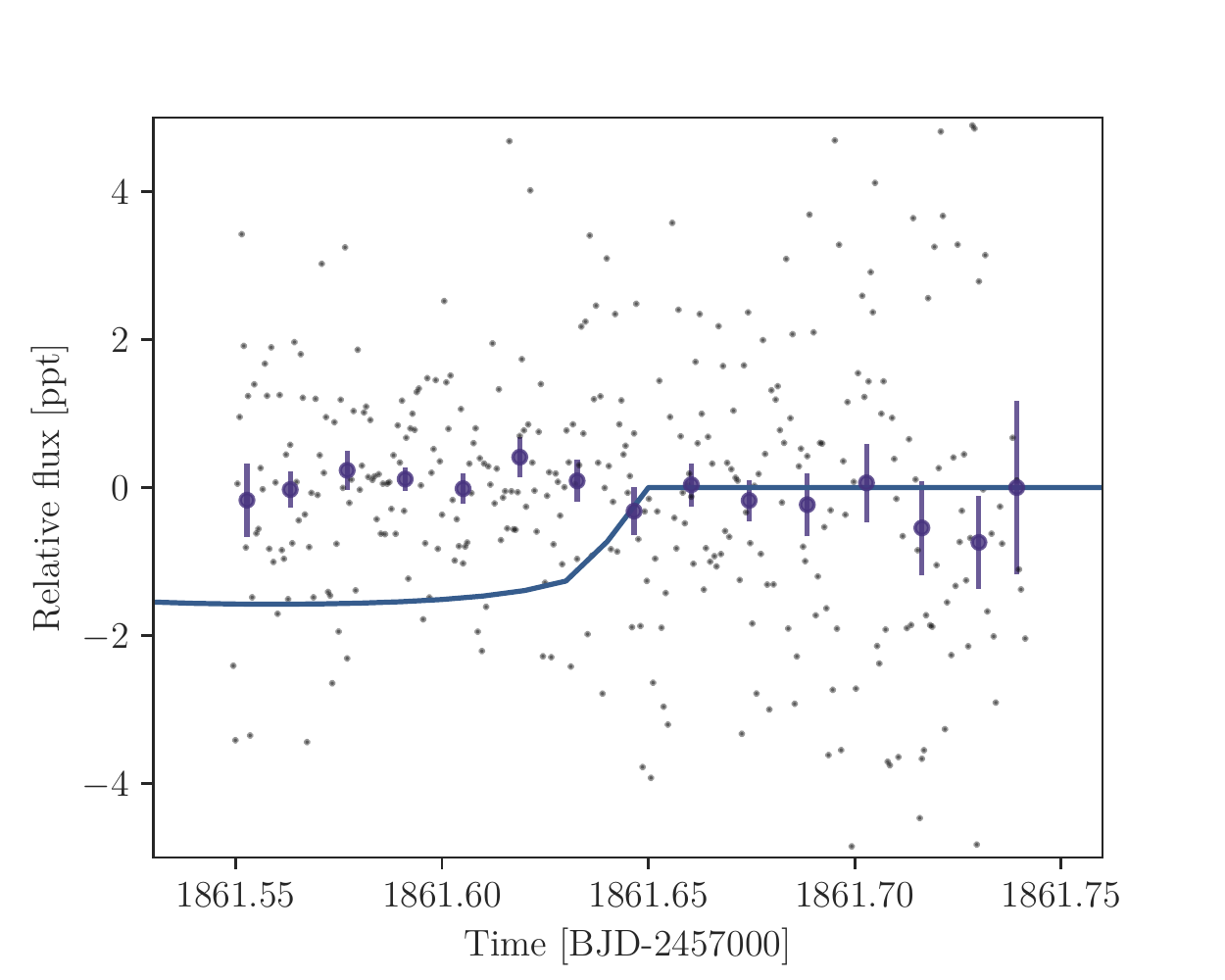}
    \caption{LCOGT photometry from the 1-m telescope at McDonald Observatory in 2020 which ruled out the 11.767d period initially detected for TOI-1471.01. Grey points show individual points, purple circles show binned photometry (and errors), and the blue line represents the expected transit model given the initial TOI data.}
    \label{fig:LCO}
\end{figure}

\begin{figure}
	\includegraphics[width=\columnwidth]{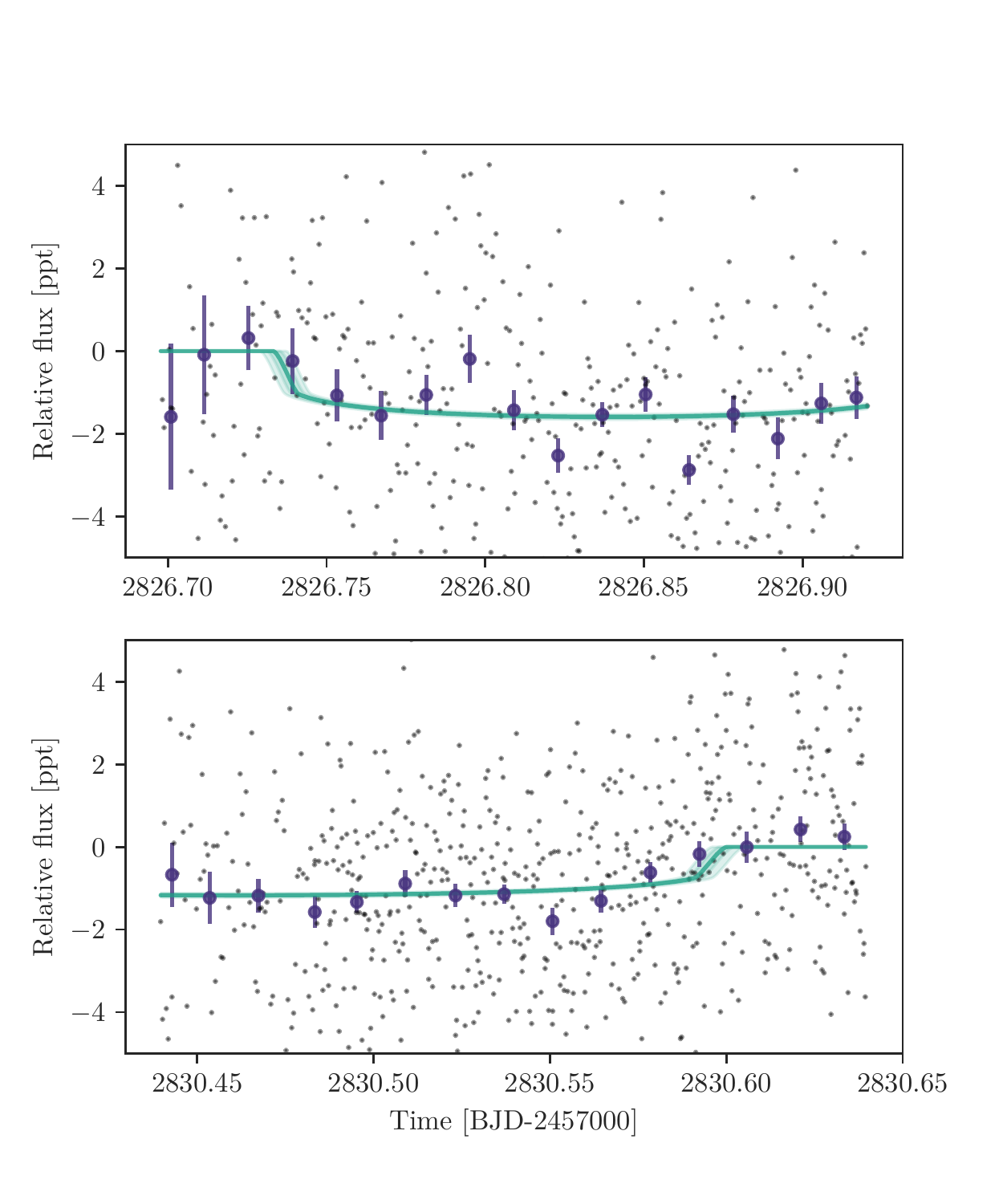}
    \caption{2022 observations of \Tstar{} using LCOGT 1-m telescopes at CTIO and SAAO respectively which confirmed the orbital periods proposed here. \refcom{\textbf{Upper}: Ingress of \Tstar{}\,b; \textbf{Lower}: Egress of \Tstar{}\,c}. Grey points show individual points, circles show binned photometry (and errors), and the green line represents the expected transit model given priors from our final combined model.}
    \label{fig:LCO2}
\end{figure}

In the case of the 2020 observation, the LCOGT light curve ruled out the expected 1.6 ppt egress in the TOI-1471 aperture as shown in Figure \ref{fig:LCO}. The LCOGT data also ruled out egress events deep enough to cause the signal in the TESS data in all four other Gaia neighbors within $2\farcm5$ of TOI-1471 that are bright enough to be capable of causing the TESS detection. The combination of results ruled out the initial TOI-1471.01 orbital period of 11.767~d, which prompted us to further investigate the events in the TESS data. 
For the 2022 observations, a simple transit model combined with decorrelation for simple metadata such as airmass \& PSF FWHM reveals strong evidence for transits at the purported orbits (see Figure \ref{fig:LCO2}).

\subsection{CHEOPS}\label{sect:Cheops} 
\cheops{} is a 30-cm ESA space telescope devoted to the characterisation of exoplanets from its position in a sun-synchronous low-Earth orbit \citep{Benz2021}.
\Tstar{} was observed on four occasions through \cheops{} GTO program \#48 "Duos: Recovering long period duo-transiting planets with \cheops{}" in an attempt to recover the true period of \Tstar{}\,c.
Four visits were scheduled (see Table \ref{tab:cheops}), each with a duration of 7.1 \cheops{} orbits (11.85hr) and an exposure time of 46.65s.
The data for the first three \cheops{} visits is shown in Figure \ref{fig:CheopsOther}.
We used the \texttt{make\_xml\_files} function of \texttt{pycheops} \citep{maxted2022}
to generate the visits and exposure time.

\begin{table}
    \caption{Key information for the \cheops{} photometry presented in this paper. The raw and detrended \cheops{} photometry shown in this paper is available on Vizier.}
    \label{tab:cheops}
    \tiny{
    \begin{tabular}{lccc}
        \hline
        \hline
        Start time [UT \& BJD] & Dur [hrs] & Aliases [d] & File ref. \\
        \hline
2021/11/24 04:37:17 \/ 2459542.69256 & 11.501 & 40.20d & CH\_PR110048\_TG017601\_V0200 \\
2021/12/04 20:39:18 \/ 2459553.36063 & 11.578 & 45.56 \& 22.78d & CH\_PR110048\_TG017401\_V0200 \\
2021/12/11 08:47:17 \/ 2459559.86617 & 10.515 & 97.62, 48.81 \& 24.40 &  CH\_PR110048\_TG017301\_V0200 \\
2021/12/18 22:42:58 \/ 2459567.44651 & 11.578 & 52.56d & CH\_PR110048\_TG017201\_V0200 \\
    \hline
    \end{tabular}
     }
\end{table}

\begin{figure}
    \centering
    \includegraphics[width=\columnwidth,trim={8 8 8 8}]{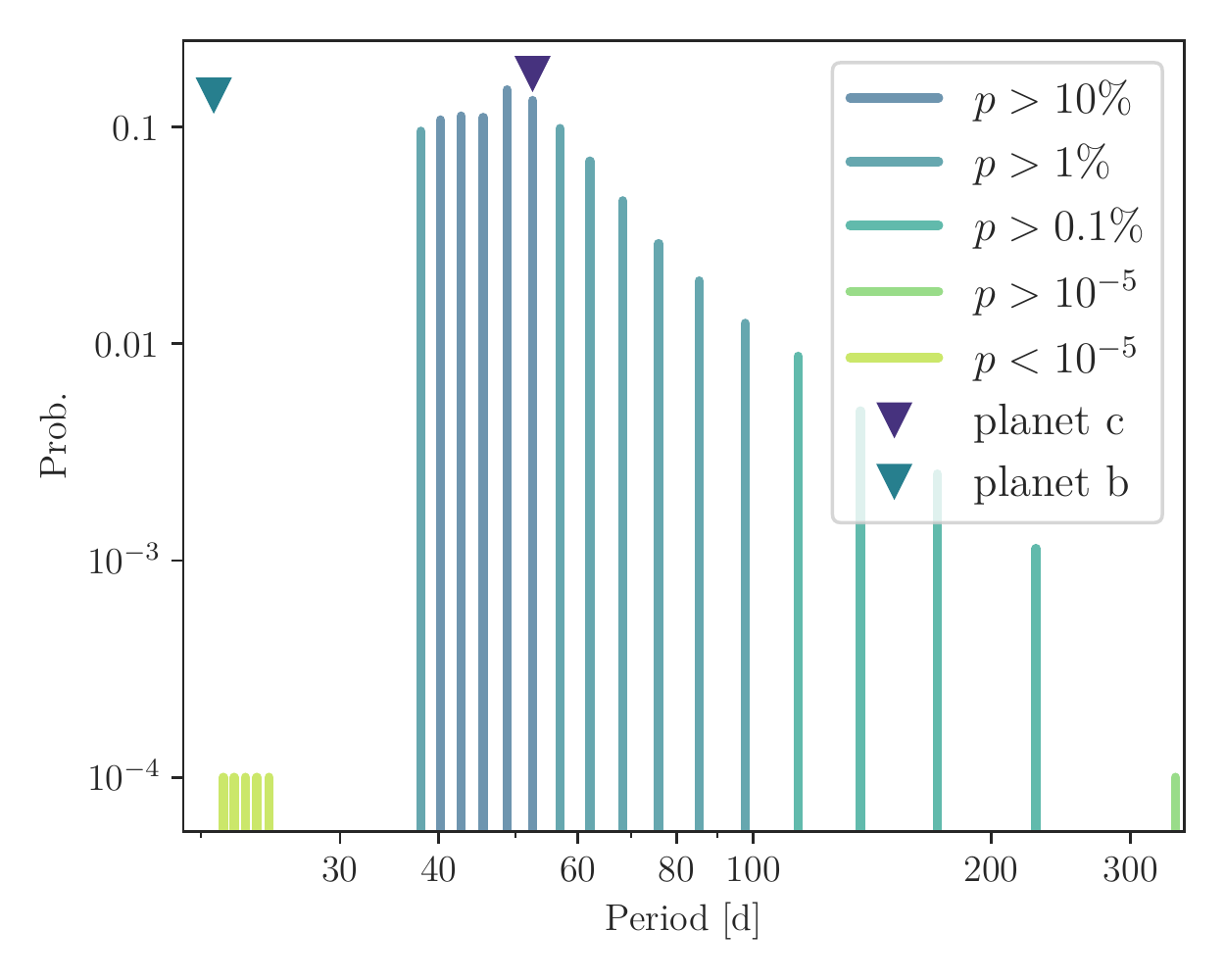}
    \caption{Marginalised probabilities for each of the period aliases of \Tstar{}\,c as calculated by \texttt{MonoTools}. Green bars for the period aliases $<30$d are simply representative - these aliases actually have probabilities $<10^{-45}$.}
    \label{fig:tessonly}
\end{figure}

\subsubsection{PIPE} 
We extracted photometry from the CHEOPS data by modelling the Point Spread Function (PSF) using the custom ``PSF imagette photometric extraction'' (\texttt{PIPE}) package  \footnote{\url{https://github.com/alphapsa/PIPE}}, which has also been used in past \textit{CHEOPS} analyses \citep[e.g.][]{Szabo2022,Serrano2022}.
This uses the smaller but shorter-cadence imagettes and a measured PSF template to measure the underlying stellar flux variations. 
Comparison with the CHEOPS DRP \citep{Hoyer2020} reveals similar photometric precision, but with less severe trends as a function of e.g. roll angle. Hence we chose to use the PIPE photometry.
\refcom{PIPE is also able to use a map of detector hot pixels to remove noise due to variable hot pixels in the region of the PSF.
It also applies a non-linearity correction to the background which reduces the number of outliers.}
To model the \Tstar{} imagettes we used a PSF model derived for HD\,209458, which differs in \teff{} by only 400\,K and was observed at the same pixel location.

\subsubsection{Analysis of period aliases} \label{sect:TESS_only_analysis}
In order to assess the ability of \cheops{} to observe period aliases of \Tstar{}\,c, we performed an initial analysis using \texttt{MonoTools} \citep{Osborn2022b} with the \tess{} data alone.
This is an open-source transit-fitting package designed specifically for planets on long period orbits which produce only one or two transits. 
In the case of a so-called \textit{duotransit} like \Tstar{}\,c - i.e. a transiting planet candidate with two observed transit but multiple possible period aliases - \texttt{MonoTools} takes care to calculate a precise period probability distribution.

The fitting process for duotransits is further described in \citet{Osborn2022}, but we briefly explain it here.
First, we simultaneously fit the inner periodic planet, and the two transits of the outer planet using the \texttt{exoplanet} package \citep{exoplanet:joss} - this ensures that any constraints from the inner planet on e.g. stellar density can help further constrain the outer body.
For the duotransit candidate, we use the available \tess{} photometry to calculate the possible array of unobserved period aliases. In the case of \Tstar{}\,c there are 23 possible solutions. 
Unlike the inner planet, the duotransit is fitted using the transit shape alone (i.e. a way that is agnostic of orbital period) as well as using the central time of transit for each of the two transits.
The resulting shape parameters constrain the transverse velocity of \Tstar{}\,c across the star.

To assess the marginalised probability across all period aliases, we can use all available information to produce constraining priors. 
First, we consider a simple period prior which incorporates the window function (i.e., given some observation $\tau$, the probability of observing transits during that window decreases with $\tau/P$, i.e. $p\propto P^{-1}$) and a factor to account for the fact that planetary occurrence is roughly uniform in $\log{P}$ rather than P, again meaning $p\propto P^{-1}$.
These are detailed in \citet{kipping2018orbital}.
Next we include a geometric prior - i.e. transit probability is inversely proportional to the distance during transit: $p\propto{a_0/R_s}$.

We also have knowledge of the expected eccentricity of \Tstar{}\,c, both due to internal constraints within the planetary system (i.e. we know that its orbit cannot graze its host star or cross the orbit of \Tstar{}\,b), and also from the average distributions of exoplanetary eccentricities.
For the former maximum eccentricity, we use the hill sphere of \Tstar{}\,b (assuming the average mass for its radius) to compute a maximum eccentricity for each potential period alias.
For the latter population distribution, we use the eccentricity distribution of \citet{van2019orbital} to constrain the expected eccentricity of \Tstar{}\,c.
Due to the form of the eccentricity distributions and the distribution of possible eccentricities ($e$), argument of periasteron ($\omega$) for each transverse velocity ($v/v_{\rm circ}$), there is no analytical solution for this calculation, therefore we used $1.3\times10^{8}$ samples in $v/v_{\rm circ}$ and $\omega$ space to create 2D interpolated distributions of the prior probability as a function of both $v/v_{\rm circ}$ and $e_{\rm max}$.

We then combine all priors for each period alias and normalise the sum of the combined probabilities to 1, allowing us to assign marginal probabilities to each alias.
We then sampled the fit using the Hamiltonian Monte Carlo sampler implemented in \texttt{PyMC3} \citep{exoplanet:pymc3} for 6000 steps across 6 chains, producing thousands of independent and well-mixed samples ($\hat{r}$<1.05).

The result can be seen in Figure \ref{fig:tessonly}.
Due to our eccentricity prior eliminating aliases with orbits that are highly likely to cross that of \Tstar{}\,b, the five aliases between 21.35 \& 24.40d were excluded.
Instead, the model showed a clear peak on periods of 40-60d, with the average period across all aliases weighted by probability being $52.5\pm14.6$d.
We also inspected Keck/HIRES and Lick/APF RVs taken by the TKS group, however these did not prove constraining on the orbit of \Tstar{}\,c, and will be published in a future analysis.
This led us to schedule 9 aliases between 38 \& 68d on \cheops{}.
Of these, three were priority 1 aliases, which had the highest chance of being observed - 40.2, 45.6 \& 52.5d aliases. 
The first was to verify if the system was near a 2:1 period ratio, the second was due to a marginal peak in the Keck/HIRES and Lick/APF RVs, and the third was due to this being close to the peak in our \texttt{MonoTools} solution.

\begin{figure}
	\includegraphics[width=\columnwidth, trim={5 10 10 10}]{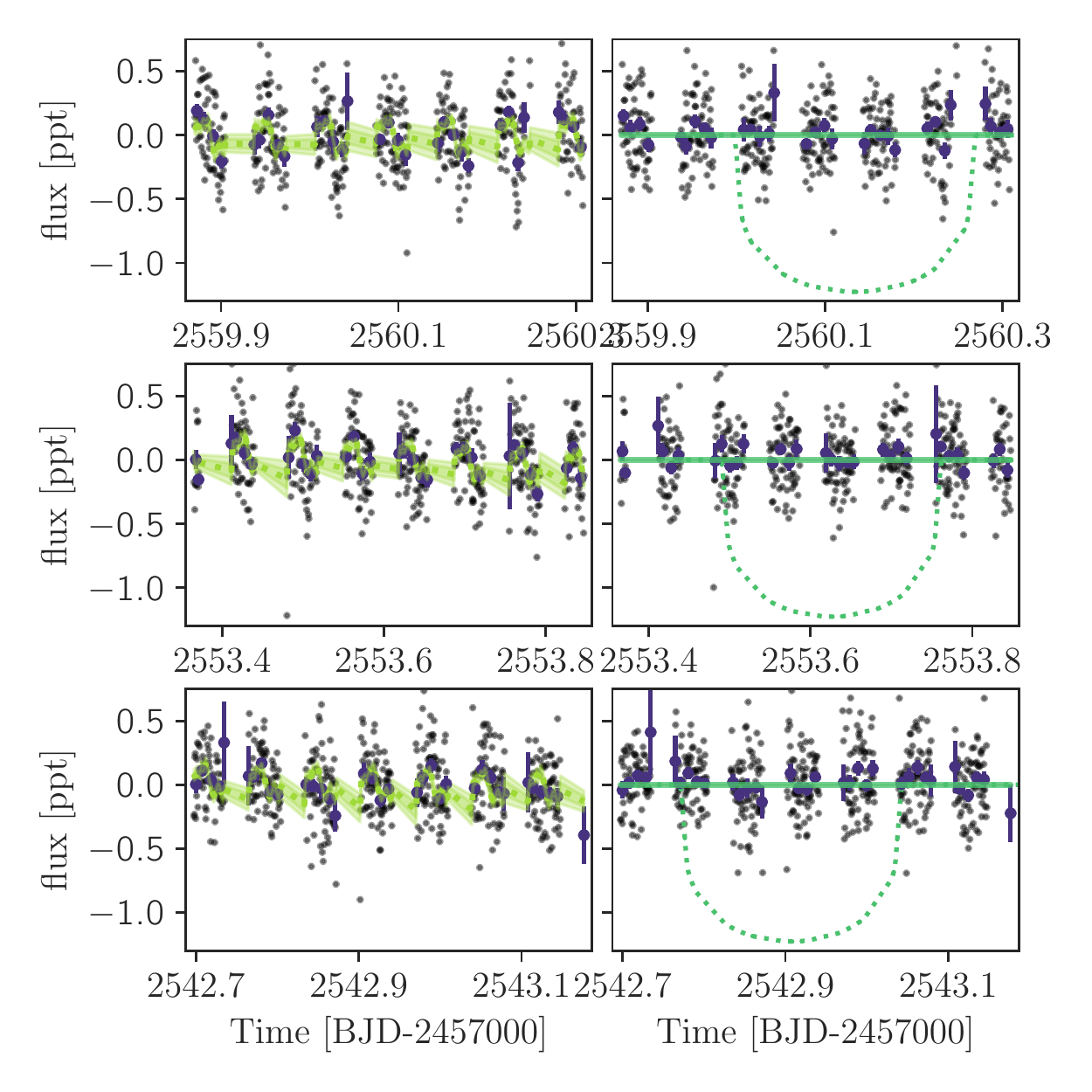}
    \caption{\cheops{} photometry for the three early \& unsuccessful visits of \Tstar{}. Left-hand plots show the raw PSF photometry, as well as best-fit regions for a simple decorrelation model (green). Right-hand plots show the residual photometry, which together has an RMS of only $8{\rm ppm}\,{\rm hr}^{-1}$. \refcom{In both plots, 15-minute flux bins are shown in purple.} The expected transit model of \Tstar{}\,c is \refcom{shown as a dashed line}.}
    \label{fig:CheopsOther}
\end{figure}
\begin{figure}
	\includegraphics[width=\columnwidth]{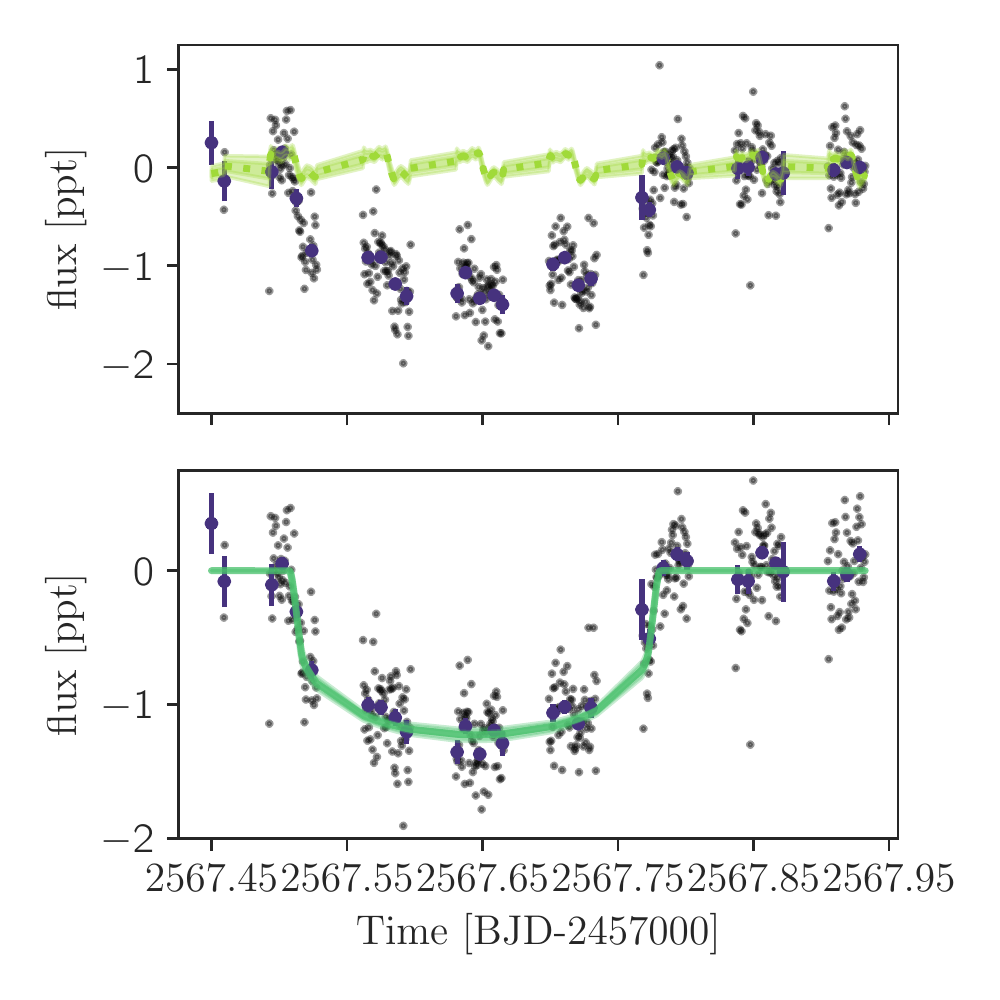}
    \caption{\cheops{} photometry of the visit at 52.5d. \cheops{} data is represented by grey individual points and purple binned flux \& errors. \refcom{The upper figure shows} the raw PIPE photometry with the detrending model $1$ \& $2\sigma$ error regions (in two transparency steps) and best-fit (dashed line) in green. \refcom{The lower figure shows} detrending-removed \cheops{} photometry and $1$ \& $2\sigma$ error regions and best-fit (dashed line) for the \Tstar{}\,c transit model in turquoise.}
    \label{fig:CheopsTransit}
\end{figure}

After three unsuccessful observations, (see Figure \ref{fig:CheopsOther}), we successfully recovered a transit of \Tstar{},c with \cheops{} on the fourth attempt, shown in Figure \ref{fig:CheopsTransit}.
As with previous cases \citep[e.g.][]{Osborn2022}, this was close to the maximum suggested by \monotools{} modelling, helping to once again validate our modelling approach.

\subsection{Spectra}
\subsubsection{FIES}\label{sect:FIES} 
We used the FIbre-fed Échelle Spectrograph \citep[FIES;][]{Frandsen1999,Telting2014} mounted on the 2.56 m Nordic Optical Telescope (NOT) of Roque de los Muchachos Observatory (La Palma, Spain) to acquire 5 high-resolution ($R\sim67000$) spectra of \Tstar{} over 15\,d.
We used "Stellar parameter classification" \citep[SPC;][]{Buchhave2012,Buchhave2014} producing stellar parameters of \teff{}$=5638\pm55$~K, \logg{}$=4.44\pm0.10$, $[\rm{m/H}]=-0.04\pm0.08$, $v\sin{i}=2.2\pm0.5$~\kms{}, all of which are within the uncertainties of our final derived values presented in Section \ref{sect:starpars} and Table \ref{table:star_params}.
The FIES RV measurements show no significant variation within the measurement uncertainties, however they remain too low-precision $\sigma_{\rm RV}>150$~\ms{} to prove useful in detecting the reflex motion of the two planets.

\subsubsection{TRES}\label{sect:TRES} 
Four reconnaissance spectra were obtained between December 2019 and February 2020 using the Tillinghast Reflector Echelle Spectrograph \citep[TRES;][]{gaborthesis} mounted on the 1.5m Tillinghast Reflector telescope at the Fred Lawrence Whipple Observatory (FLWO) atop Mount Hopkins, Arizona. TRES is a fiber-fed echelle spectrograph with a wavelength range of 390-910nm and a resolving power of R$\sim44,000$. The spectra were extracted as described in \citet{Buchhave2010} and then used to derive stellar parameters using the Stellar Parameter Classification tool \citep[SPC;][]{Buchhave2012}. SPC cross-correlates an observed spectrum against a grid of synthetic spectra derived from Kurucz atmospheric models \citep{Kurucz1992}. The averaged stellar parameters (\teff = $5637\pm50$ K, \logg = $4.40\pm0.10$, \meh = $-0.08\pm0.08$, and \vsini = $2.5\pm0.5$~\kms) agree well with the adopted parameters. 

\subsubsection{HARPS-N}\label{sect:HARPS-N}
Between January 20$^\mathrm{th}$ 2021 (UT) and January 30$^\mathrm{th}$ 2022 (UT) we collected 13 spectra with the High Accuracy Radial velocity Planet Searcher for the Northern hemisphere \citep[HARPS-N: $\lambda$\,$\in$\,(378--691)\,nm, R$\approx$115\,000,][]{2012SPIE.8446E..1VC} mounted at the 3.58-m Telescopio Nazionale Galileo (TNG) of Roque de los Muchachos Observatory in La Palma, Spain, under the observing programs ITP19\_1 (see Table \ref{tab:harpsn_rvs}).
The exposure time was set to 280--2400 seconds, based on weather conditions and scheduling constraints leading to a SNR per pixel of 28--107 at 550\,nm. The spectra were extracted using the off-line version of the HARPS-N DRS pipeline \citep{2014SPIE.9147E..8CC}, version {\tt HARPN\_3.7}.
Absolute radial velocities (RVs) and spectral activity indicators (CCF\_FWHM, CCF\_CTR, BIS and Mont-Wilson S-index) were measured on the higher-preicison \harpsn{} spectra using an on-line version of the DRS, the YABI tool\footnote{Available at \url{http://ia2-harps.oats.inaf.it:8000}.}, by cross-correlating the extracted spectra with a G2 mask \citep{1996A&AS..119..373B}.
We also used {\tt serval} code \citep{zechmeister2018spectrum} to measure relative RVs by the template-matching, chromatic index (CRX), differential line width (dLW), and H$\alpha$, and sodium Na~D1 \& Na~D2 indexes. The uncertainties of the relatives RVs measured with {\tt serval} are in the range 0.7--3.1\,\ms, with a mean value of 1.5\,\ms.
The uncertainties of absolute RVs measured with the online version of DRS (YABI) are in a the range 0.8--4.2\,\ms, with a mean value of 1.6\,\ms.
Table~\ref{tab:harpsn_rvs} gives the time stamps of the spectra in BJD$_{\mathrm{TDB}}$, {\tt serval} relative RVs along with their $1\sigma$ error bars, and spectral activity indicators measured with YABI and {\tt serval}.
In the joint RV and transit analysis presented in Section \ref{sect:model} we used relative RVs measured from HARPS-N spectra with {\tt serval} by the template-matching technique.

\subsubsection{SOPHIE}\label{sect:Sophie} 
\sophie{} is a stabilized \'echelle spectrograph dedicated to high-precision RV measurements on the 193-cm Telescope at the Observatoire de Haute-Provence, France \citep{Perruchot2008}. Before having been identified as TOI-1471, \Tstar{} was first observed in 2011 with \sophie{} as part of its volume-limited survey of giant extrasolar planets \citep[e.g.][]{Bouchy2009,Hebrard2016}. After its identification as the host of the transiting planet candidate TOI-1471.01, it was re-observed in 2019-2021 with \sophie{} as part of its TESS follow-up. Overall, we secured 28 \sophie{} spectra of \Tstar{} in its high resolution mode (resolving power $R=75 000$). Depending on the weather conditions, their exposure times ranged from 3 to 20 minutes (typically 10 minutes) and their signal-to-noise ratio per pixel at 550~nm from 30 to 80. The corresponding radial velocities were extracted with the standard \sophie{} pipeline using cross correlation functions 
\citep{Bouchy2009} and including CCD charge transfer inefficiency correction \citep{Bouchy2013}. Following the method described in \citet{Pollacco2008} and \citet{Hebrard2008}, we estimated and corrected for the moonlight contamination using the second \sophie{} fiber aperture, which is targeted on the sky while the first aperture points toward the star. 
We estimated that only three spectra were significantly polluted by moonlight; one of them was too contaminated and was discarded, whereas the other two were corrected, with a correction value below 10\,\ms{}.

Thus our final \sophie{} dataset included 27 measurements showing uncertainties ranging from 1.6 to 3.9~\ms{} (see Table \ref{tab:sophie_rvs}). The RVs show a 9-\ms{} dispersion around a blueshifting drift of about 200~\ms{} in ten years. The corresponding bisectors of the cross correlation functions do not show any significant variation nor correlation with the RV, so there are no hints for RV variations induced by blend configurations nor stellar activity.

\subsubsection{CAFE}\label{sect:cafe} 
Finally, we also observed TOI-1471 with the \cafe{} (Calar Alto Fiber-fed Echelle) spectrograph \citep{aceituno13} mounted at the 2.2m telescope of the Calar Alto observatory. A total of 22 spectra were obtained between 18-Dec-2019 and 19-Aug-2022 with typical signal-to-noise ratio of 30 (see Table \ref{tab:cafe_rvs}). The data were reduced and the spectra extracted by using the observatory pipeline described in \cite{lillo-box19}, which also determines the radial velocity by performing cross-correlation against a solar binary mask. Usually, several spectra were obtained for each night. We binned the RVs per night. This implies a total of 10 measurements. We discarded the first one due to lack of radial velocity standards observed that night that prevented us from calculating and correcting for the relevant nightly zero point. Hence, nine measurements are available with a median uncertainty of 7 \ms{}.

\subsection{High-Resolution Imaging}
A detected exoplanet transit signal might be a false positive due to a background eclipsing binary or yield incorrect stellar and exoplanet parameters if a close companion exists and is unaccounted for \citep{FH1, FH2}.
Additionally, the presence of a close companion star leads to the non-detection of small planets residing with the same exoplanetary system \citep{Lester}.
Given that nearly one-half of FGK stars are in binary or multiple star systems \citep{Matson}, high-resolution imaging provides crucial information toward our understanding of exoplanetary formation, dynamics and evolution \citep{Howell2021}.

As part of the standard process for validating TESS candidates \citep[e.g.][]{ciardi2015}, we observed \Tstar{} with a combination of high-resolution resources including near-infrared adaptive optics (AO) imaging at Palomar and Keck and optical speckle observations at Gemini-North, SOAR, and SAI-2.5m.  While the optical observations tend to provide higher resolution, the NIR AO tend to provide better sensitivity, especially to lower-mass stars. The combination of the observations in multiple filters enables better characterization for any companions that may be detected.  Gaia DR3 is also used to provide additional constraints on the presence of undetected stellar companions as well as wide companions.

\begin{table}
    \caption{Key information for high-resolution Imaging.}
    \label{tab:hires}
    \begin{tabular}{lcccc}
        \hline
        \hline
        Facility & Instrument & Filter & Image time [UT \/ BJD]\\
        \hline
        Keck2 & NIRC2 & BrGamma & 2020-05-28 \/ 2458997.82 \\
        SOAR & HRCam & I & 2020-10-31 \/ 2459153.65 \\
        SAI-2.5m & Speckle Polarimeter & 625nm & 2020-12-03 \/ 2459186.25 \\ 
        Gemini & 'Alopeke & 562nm & 2020-12-04 \/ 2459187.75 \\
        Gemini & 'Alopeke & 832nm & 2020-12-04 \/ 2459187.75 \\
        Palomar & PHARO & BrGamma & 2020-12-04 \/ 2459187.75 \\
        Gemini & 'Alopeke & 562nm & 2021-10-17 \/ 2459504.8 \\
        Gemini & 'Alopeke & 832nm & 2021-10-17 \/ 2459504.8 \\
        \hline
    \end{tabular}
\end{table}

Detailed descriptions for the observations, reduction \& analysis for all five high-resolution instruments are found in sections \ref{sect:Keck_palomar}, \ref{sect:Gemini}, \ref{sect:SOAR} \& \ref{sect:SAI} and summarised in Table \ref{tab:hires}.
In summary, none of the observations revealed evidence for close-in ($\lesssim 1\arcsec$) stellar companions.
However, the Palomar observations did detect a faint stellar companion 7 magnitudes fainter than the primary star and 3\arcsec\ to the southwest (PA=235$^\circ$).  The companion star is Gaia DR3 94468978202368768.

\subsection{Gaia \& Archival Assessment}\label{sect:gaia}
In addition to the high resolution imaging, we have used Gaia to identify any wide stellar companions that may be bound members of the system.  Typically, these stars are already in the TESS Input Catalog and their flux dilution to the transit has already been accounted for in the transit fits and associated derived parameters.  There are no additional widely separated companions identified by Gaia that have the same distance and proper motion as \Tstar{} \citep[see also ][]{mugrauer2020,mugrauer2021}.
    
The faint stellar companion detected by Palomar was detected by Gaia (DR3 94468978202368768) but a full astrometric solution (parallax and proper motion) for this star is not yet available in DR3.  However, the epoch of observations between Gaia (2016.0) and Palomar (2020.93) are sufficiently separate to use the proper motion as a test for boundedness. \Tstar{} has a proper motion of $\mu_{\alpha} = 158.8$ mas/yr and $\mu_{\delta} = 107.5$ mas yr$^{-1}$, which should produce an increase in separation of the two stars of $\Delta\alpha = 0.78\arcsec$ and $\Delta\delta = 0.53\arcsec$ for a total separation increase of $0.95\arcsec$.  The measured separation of the stars by Gaia (2016) is 2.1\arcsec\ and by Palomar (2020) is 3.1\arcsec\ - fully consistent with the measured proper motion of TOI~1471 and the companion star being a low-proper motion background star.  Thus, the detected companion is almost certainly unbound and unrelated to the \Tstar{} system.
    
Additionally, the Gaia DR3 astrometry provides additional information on the possibility of inner companions that may have gone undetected by either Gaia or the high resolution imaging. The Gaia Renormalised Unit Weight Error (RUWE) is a metric, similar to a reduced chi-square, where values that are $\lesssim 1.4$  indicate that the Gaia astrometric solution is consistent with the star being single whereas RUWE values $\gtrsim 1.4$ may indicate an astrometric excess noise, possibily caused the presence of an unseen companion \citep[e.g., ][]{ziegler2020}.  TOI~1471 has a Gaia DR3 RUWE value of 1.02 indicating that the astrometric fits are consistent with the single star model.   

Due to the large proper motion of \Tstar{}, we were also able to use archival photometric plates\footnote{Through the APPLAUSE project; \url{https://www.plate-archive.org/query/}} to assess the possibility of a coincident bright background star. 
These observations provide a baseline of over 100 years and therefore a relative offset of 25\arcsec{}, and find no sign of any bright source.
POSS-I imaging from 1954 does detect the faint Gaia source (DR3 94468978202368768) 11\arcsec{} from \Tstar{}, confirming that it is not a bound companion.

\section{Analysis}

\subsection{Stellar parameters} \label{sect:starpars} 
The stellar parameters for \Tstar{} are presented in Table \ref{table:star_params}. The analysis of the co-added HARPS-N stellar spectrum has been carried out by using the \texttt{BACCHUS} code \citep{2016ascl.soft05004M} relying on the MARCS model atmospheres \citep{2008A&A...486..951G} and atomic and molecular line lists from \citet{2021A&A...645A.106H}. In brief, the surface gravity ($\rm \log g = 4.47 \pm 0.10$) has been determined by requiring ionization balance of \ion{Fe}{I} lines and \ion{Fe}{II} line. A microturbulence velocity  has also been derived ($\rm 0.95 \pm 0.1$\kms{}) by requiring no trend of Fe line abundances against their equivalent widths. The output metallicity  ($\rm [Fe/H] = -0.07 \pm 0.06$) is represented by the average abundance of the \ion{Fe}{I} lines. An effective temperature of 5611\,$\pm$\,31\,K has been derived  by requiring no trend of the \ion{Fe}{I} lines abundances against their respective excitation potential.

We used the \harpsn{} spectra to measure the stellar rotation ($v \sin i$) using the average of the Fe lines broadening after having subtracted the instrument and natural broadening. This technique led to an upper limit of the stellar rotation velocity of $<$3.5\kms{}, which agrees well with the $2.5\pm0.5$\kms{} derived from FIES spectra, both of which highlight a long stellar rotation period. 

In a second step, we used the Bayesian tool PARAM \citep{2014MNRAS.445.2758R,2017MNRAS.467.1433R} to derive the stellar mass, radius, and age utilizing the spectroscopic parameters and the updated \textit{Gaia} luminosity along with our spectroscopic temperature. However, such bayesian tools underestimate the error budget as they do not take into account the systematic errors between a set of isochrones to another due to the various underlying assumptions in the respective stellar evolutionary codes. In order to take into account those systematic errors, we combined the results of the two sets of isochrones provided by PARAM (i.e. MESA and Parsec) and add the difference between the two sets of results to the error budget provided by PARAM. We obtained a stellar radius and mass of  respectively $\rm 0.97 \pm 0.02\,R_\odot$ and $\rm 0.89 \pm 0.07\,M_\odot$).  Despite its nearly solar metallicity , the derived age from the isochrones indicates that the star is old (9 $\pm$ 4 Gyr). This is consistent with the fact that we do not detect any lithium in the atmosphere of the star, nor we detect a chromospheric activity in the core of the H \& K Ca lines and that we find indications of a low rotation period.

However, we emphasize that although using two sets of isochrones may mitigate underlying systematic errors, our formal error budget for radius and luminosity may still be underestimated, as demonstrated by \citet{Tayar2022}. For solar-type stars such as TOI-1471, absolute errors may rather be up to 4\%, 2\%, 5\% and 20\% for respectively radius, luminosity, mass and age.

\subsubsection{Rotation}
Based on \citet{1984ApJ...279..763N} and \citet{2008ApJ...687.1264M} activity-rotation relations and using (B-V) of 0.688 and the \logrhk{} measured with YABI (-4.972\,$\pm$\,0.044), we estimated a rotation period of \Tstar{} of 29.0\,$\pm$\,5.9\,days and 30.4\,$\pm$\,3.5\,days, respectively.
This is consistent with the \vsini\ upperlimits derived from spectra.
Using the activity-age relation of \citet{2008ApJ...687.1264M} we also found an age of \Tstar{} to be in a range of 3.6--7.3\,Gyr, consistently with the age determined with the isochrones within the errorbars.

\subsubsection{Radius}
We employed a Markov-Chain Monte Carlo (MCMC) modified infrared flux method (IRFM; \citealt{Blackwell1977,Schanche2020}) to determine the radius of \Tstar{} by computing the stellar bolometric flux and obtaining the effective temperature and angular diameter. This was done by fitting spectral energy distributions (SEDs), constructed using priors from our spectral analysis, to broadband fluxes and uncertainties from the most recent data releases for the following bandpasses; {\it Gaia} G, G$_{\rm BP}$, and G$_{\rm RP}$, 2MASS J, H, and K, and {\it WISE} W1 and W2 \citep{Skrutskie2006,Wright2010,GaiaCollaboration2021}. The angular diameter is converted to a radius using the offset-corrected {\it Gaia} parallax \citep{Lindegren2021}. To account for stellar model biases we conducted a Bayesian modelling averaging of the \textsc{atlas} \citep{Kurucz1993,Castelli2003} and \textsc{phoenix} \citep{Allard2014} catalogues to produce weighted averaged posterior distribution of the stellar radius and obtain a value of $R_\mathrm{s}=0.962\pm0.008\, R_{\odot}$. Due to the non-luminous and faint nature of the companion we do not include a second SED and corresponding free parameters in the MCMC.


\begin{table}
\centering                          
\begin{tabular}{l c c }        
\hline\hline                 
Parameter & Value & source \\
\hline                        
Name & \Tstar{} & -- \\
TOI & \TTstar{} & -- \\
TIC & \TTIC & TICv8 \\
\textit{HD} designation & HD\,12572 & -- \\
\gaia{} DR3 ID & \TGAIA{} & Gaia DR3 \\
RA [$^\circ$, J2015.5] & \Tra{} & Gaia DR3 \\
Dec [$^\circ$, J2015.5] & \Tdec{} & Gaia DR3 \\
\tess{} mag & \TTmag{} & TICv8 \\
\textit{G} mag & \TGAIAmag{} & Gaia DR3 \\
\textit{K} mag & \TKmag{} & 2MASS \\
\teff{} [K] & \TTeff{} & This work\\
$R_s$ [$R_\odot$] & \TRs{} & This work \\
$M_s$ [$M_\odot$] & \TMs{}  & This work\\
\logg{} [cgs] & \Tlogg{} & This work\\
\hline                                   
\end{tabular}
\caption{ Stellar information for \Tstar{}.  TIC v8 described by \citep{Stassun2019}, Gaia DR3 by \citep{GaiaCollaboration2021}, Hipparcos by \citep{lindegren1997hipparcos}. 2MASS by \citep{2003tmc..book.....C}, and stellar parameters derived by our own analysis are described in Section \ref{sect:starpars}.}\label{table:star_params}    
\end{table}

\subsection{Combined photometry \& RV model}\label{sect:model}
Once we had successfully recovered the true period of \Tstar{}\,c we investigated the properties of the two transiting planets by modelling both the radial velocities and the \tess{} \& \cheops{} photometry together.
The model was built in \texttt{PyMC3} \citep{exoplanet:pymc3} using the Keplerian orbits \& transit models of \texttt{exoplanet} \citep{exoplanet:joss}. For the stellar parameters, we use the values derived in Section \ref{sect:starpars} as priors to our analysis (see Table \ref{table:star_params}). The SNR and the partial nature of the two ground-based transit observations with LCO (see Sect. \ref{sect:LCOGT}) meant we did not include them in the combined model.
 
\Tstar{} appears to be a quiet old G star without evidence of stellar activity in the \tess{} lightcurve, and thanks to the quaternion detrending systematic noise appears minimal.
Therefore we chose to remove long-duration trends using a basis spline with a knot length of 1.25\,d.
To avoid influencing the spline fit, the six transits were masked using the initially derived ephemerides.
To speed up the modelling, we then kept only 4.5 transit durations of photometry around mid-transit.
A log jitter term was included to incorporate systematic errors not incorporated into the TESS flux errors.
 
For the \cheops{} data, we performed simultaneous linear decorrelation with three normalised vectors from the \cheops{} metadata - two incorporating roll angle ($\Phi$) trends ($\sin{\Phi}$ \& $\cos{\Phi}$) and one for background flux.
We also included a log jitter term to account for excess systematic noise.
\refcom{In order to remove shorter-timescale flux variation as a function of roll angle, we included a common cubic bspline with breakpoints every $\sim8$ degrees.}
\refcom{The decorrelation terms and spline model were shared across all four \cheops{} visits.}


In section \ref{sect:companion}, we explore the possible characteristics of a massive companion in the system. This gives the possibility of an M-dwarf companion to \Tstar{} which could provide unaccounted dilution (and also depth difference between \tess{} and \cheops{} photometry).
We therefore include two dilution terms in our combined analysis using the derived expected magnitude difference of the companion in R \& I band (interpolated from primary \& secondary masses) to represent dilution in \cheops{} and \tess{} bands respectively ($9.0\pm2.1$ \& $7.4\pm1.8$).

We use informative priors on limb darkening parameters for both \tess{} and \cheops{} passbands, using the theoretical quadratic limb darkening parameters of \citet{claret2018new} \& \citet{claret2021limb}.
To account for systematic errors,  we inflate the uncertainties to 0.05 in all cases.

For the radial velocity modelling, we include a relative offset for each instrument, with priors set from the average \& standard deviation of the points.
\refcom{We also included a jitter term with a broad lognormal prior. A quadratic polynomial was used} to model the large observed drift, with normal priors with $\mu=0.0$ and $\sigma=0.05$ and $0.005$ for linear and quadratic terms.
Inspection of the activity indices from \harpsn{}, \sophie{} and \cafe{} reveals no clear rotation signal and no correlation with the radial velocities, which is consistent with the lack of obvious variability seen in the photometry.
Therefore we chose not to model activity using e.g. Gaussian Processes, and instead use only the planets Keplerian orbits.

We modelled radius ratio, $R_p/R_s$, using a broad log-normal prior to avoid unphysical negative values.
\refcom{We tested fitting RV semi-amplitude, $K_{\rm pl}$, using both broad log-normal and a free uniform prior (allowing negative values), both of which converged to consistent solutions.
We tested modelling both with and without eccentricity, testing both the \citet{VanEylen2019} multi-planet prior and the general \citet{kipping2013} priors which gave consistent results.
We use the uniform $K_{\rm pl}$ prior and \citet{kipping2013} eccentricity priors as the final model here.}

\begin{figure}
	\includegraphics[width=\textwidth, trim={15 20 15 15}]{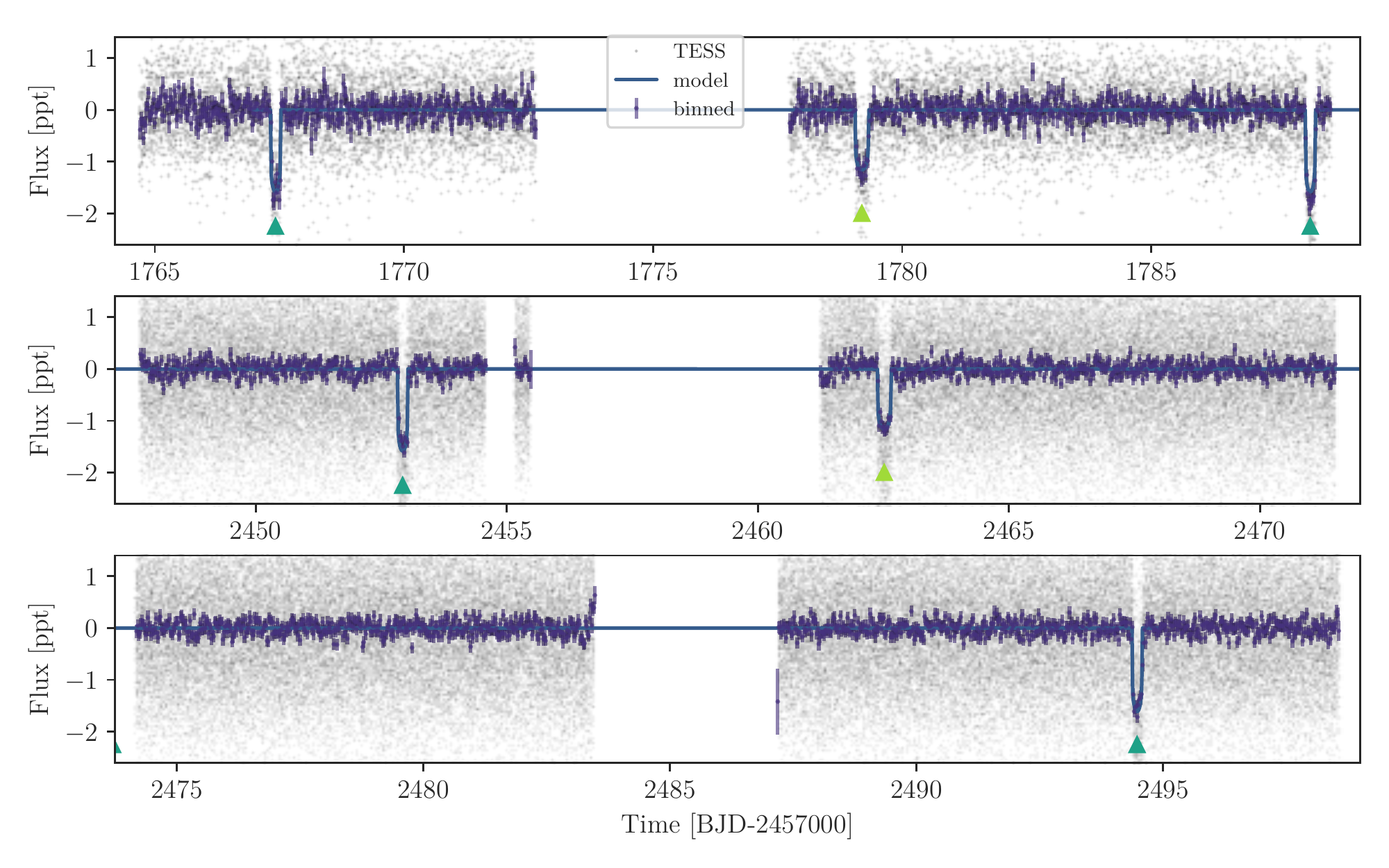}
    \caption{\tess{} photometry of \Tstar{} showing the four transits of \Tstar{}\,b and two transits of \Tstar{}\,c. Individual points are plotted as grey dots, purple circles show binned flux and errors, and the blue line corresponds the best-fit transit model.}
    \label{fig:TESS}
\end{figure}

\begin{figure}
	\includegraphics[width=\textwidth, trim={20 20 20 20}]{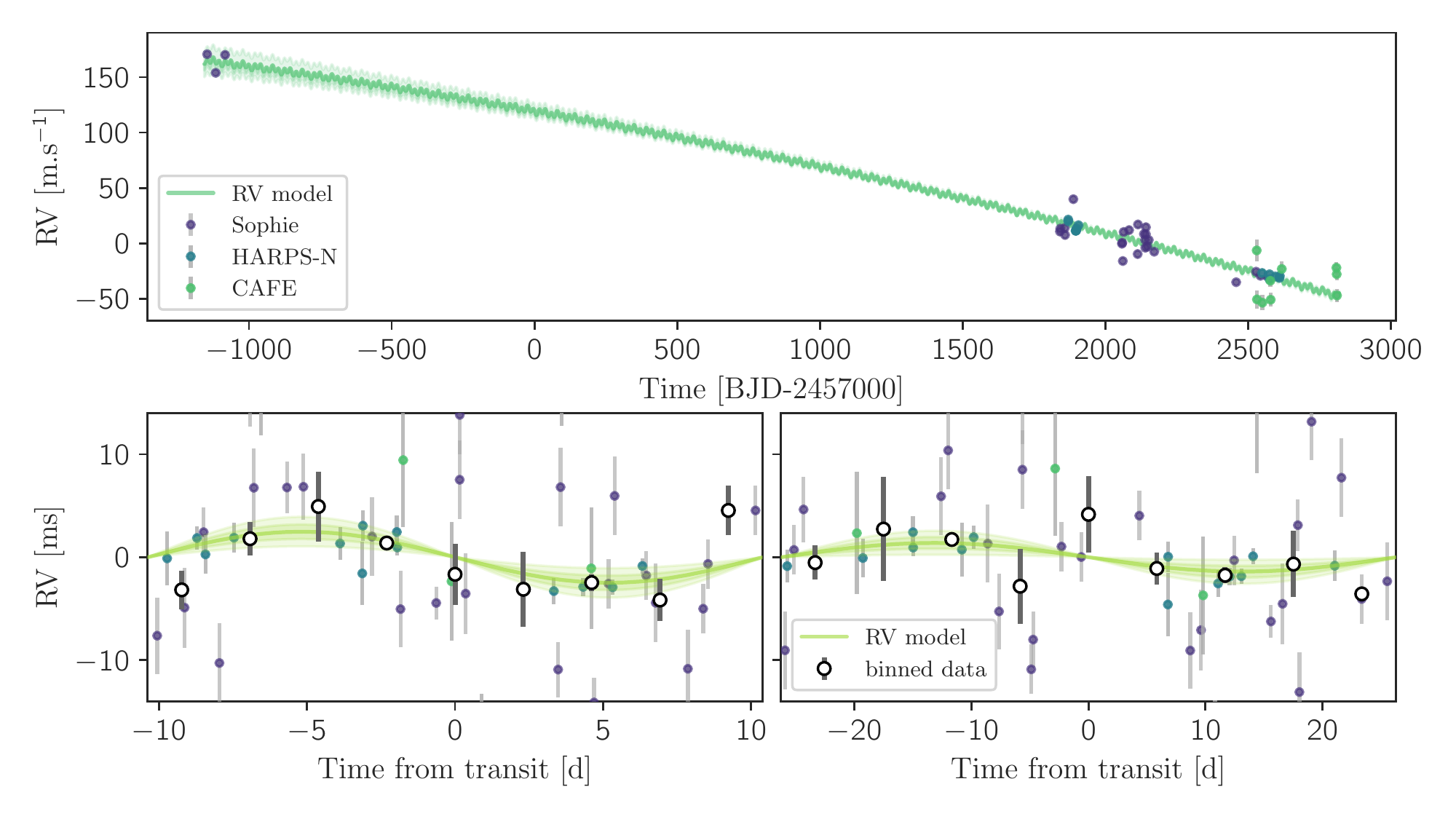}
    \caption{\sophie{}, \cafe{} \& \harpsn{} radial velocities for \Tstar{}.  Upper: RVs adjusted for intra-spectrograph offsets,  and the linear polynomial model for RV drift.  Lower panels: RVs with offsets \& trend removed showing the reflex motion from  \Tstar{}\,b (left) \& c (right). Filled areas represent $1$ and $2\sigma$ regions in two transparency steps.  \refcom{White points with black borders represent phase-binned RVs.}}
    \label{fig:RVs}
\end{figure}

The results of our combined model can be seen in Figures \ref{fig:TESS}, \ref{fig:RVs} and \ref{fig:CheopsTransit}, while the derived planetary properties can be found in Table \ref{table:planet_params}. 
All priors and posteriors for the combined modelling are shown in Table \ref{table:appendix_P35_model}

\onecolumn
\twocolumn

\begin{table*}
\centering                          
\caption{Derived planetary parameters. $^{\star}$ refers to 3-sigma upper limit (and other limits derived using this value).}             
\label{table:planet_params}      
\begin{tabular}{l c c}        
\hline\hline                 
Parameter & \Tstar{}\,b & \Tstar{}\,c \\
\hline                        
Epoch, $t_0$ [BJD-2457000] & \Ttzerozero & \Ttzeroone \\
Period, $P$ [d] & \Tperiodzero & \Tperiodone\\
Semi-major axis, $a$ [AU] & \Tsmazero & \Tsmaone  \\
Radius ratio, $R_p/R_s$ & \Trorzero & \Trorone  \\
Duration, $t_{D}$ [d] & \Ttdurzero & \Ttdurone   \\
Radius, $R_p$ [$R_\oplus$] & \Trplzero & \Trplone   \\
Insolation, $I_p$ [$Wm^{-2}$] & \TSinzero & \TSinone   \\
Surface Temp., $T_{\rm eq}$ [K] & \TTsurfpzero & \TTsurfpone   \\
RV Semi-amplitude, $K$ [\ms{}] & \TKzero & \TKone \\
Mass, $M_p/M_{\oplus}$ [\mearth{}] & \TMpzero & $<$\TMpone  \\
\refcom{Eccentricity, $e$} & \Tecczero & \Teccone \\
\refcom{Arg. of Periasteron, $\omega$} & \Tomegazero & \Tomegaone \\
TSM & \TTSMzero & \TTSMoneshort \\
\hline                                   
\end{tabular}
\end{table*}

\section{Discussion}

\subsection{RV drift}\label{sect:companion}
We measure a sizable RV drift of $-19.2$~\ms{} per year (see Figure \ref{fig:RVs}).
Comparisons of combined models with linear and quadratic RV trends also shows that the RV drift prefers a model with curvature ($-0.7\pm0.05$~\ms$\mathrm{yr}^{-2}$).
This is therefore likely indicative of a massive outer companion in the system.
Acceleration between Hipparcos and Gaia astrometric measurements is a key technique to characterising such outer companions. However, the HGAS catalogue contains only a very minimal deviation from a linear ephemeris (2.647 chi-square difference). Gaia DR3 does not flag \Tstar{} as a "non\_single\_star", though it does find a 6-sigma excess astrometric noise, however this corresponds to a dispersion of only 80~$\mu$as, and as discussed in Sect \ref{sect:gaia} and,  as discussed in Sect \ref{sect:gaia}, to a RUWE of 1.03 very likely suggests a single star.

In order to characterise the undetected outer candidate, we used the \texttt{orvara} package \citep{Brandt2021} to model the RV drift and the Hipparcos \& Gaia astrometric data jointly. 
\texttt{Orvara} accessed the Hipparcos and \gaia{} astrometry (which reveal no significant acceleration despite a 20-year baseline), and we also included our RVs with the best-fit models for planets b \& c, and the best-fit RV offsets for each instrument removed from the data.
Using our derived stellar mass as a prior, we ran 100 walkers for chains of 25 000 steps. The result, after pruning the burn-in, is a distribution of 40 000 independent samples which are consistent with the observed RV \& astrometric data.

However, we also have information derived from our high-resolution imaging observations, which are able to rule out close bright companions to \Tstar{}. 
In order to use this information, we modified \texttt{orvara} to compute the companion separation at each imaging epoch.
We used a combination of the main sequence models of \citet{Pecaut2013}\footnote{\url{http://www.pas.rochester.edu/~emamajek/EEM_dwarf_UBVIJHK_colors_Teff.txt}} and the Brown Dwarf models of \citet{Baraffe2003} to translate from companion masses derived by \texttt{orvara} into magnitudes.
We either interpolate these to the observed bands or use the closest available magnitude depending on the filter.

\begin{figure}
	\includegraphics[width=\columnwidth]{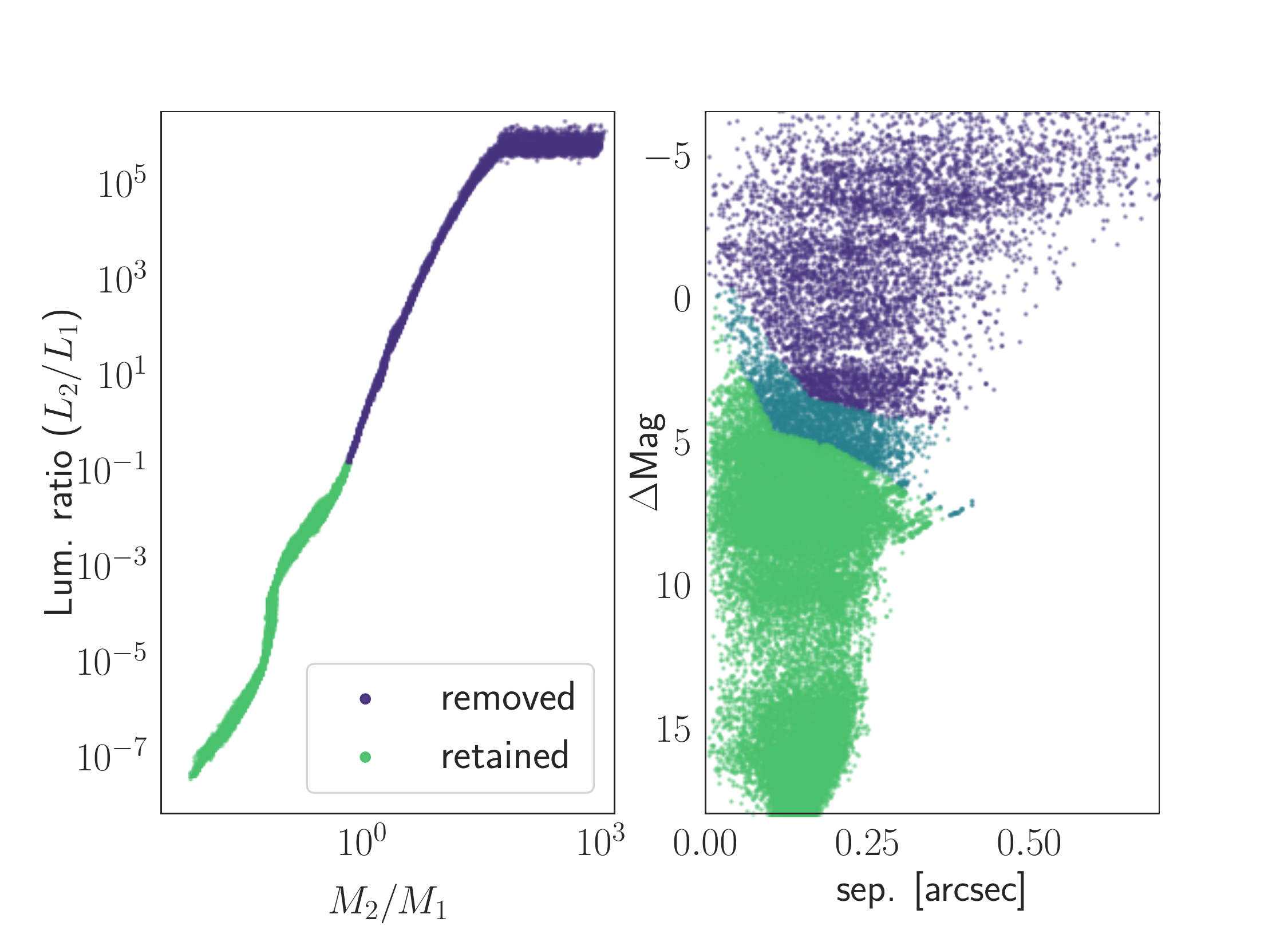}
    \caption{Samples generated with \texttt{orvara} which were accepted (in green) or rejected (in dark blue) due to a luminosity and therefore relative flux incompatible with the stacked spectrum (left), or due to a magnitude difference (or $\Delta$mag) incompatible with the multiple high-resolution images obtained in follow-up photometry. Although this pruning was performed on a per-magnitude basis, this plot shows the average magnitude across all useful bands, with samples removed due to only some imaging observations shown in blue.}
    \label{fig:orvara_prune}
\end{figure}

Given the implied separation and $\Delta$Mag, as well as an extra factor of 0.25\,mag to account for systematic noise, we were then able to compute whether our imaging observations would have detected the companion created by \texttt{orvara}.
We also have radial velocity observations and stacked spectra which should be able to resolve close-in bright binary companions. Assuming a conservative detectable flux ratio of 15\% (equivalent to a K5 star), we also removed samples which would have been detectable as contamination in e.g. the CCF.
Finally, given the inner two planets appear to be a typical multi-planet system, it is very unlikely that the low-mass companion would have extremely close encounters with the inner reaches of this system. Therefore we also included a threshold on the perihelion distance of 0.78AU - three times the orbital distance of \Tstar{}c.
This resulted in removing $\sim53\%$ of the samples generated by \texttt{orvara} -- mostly those from high-mass, large-separation and high-eccentricity regimes (see Figure \ref{fig:orvara_prune}).

\begin{figure}
	\includegraphics[width=\columnwidth]{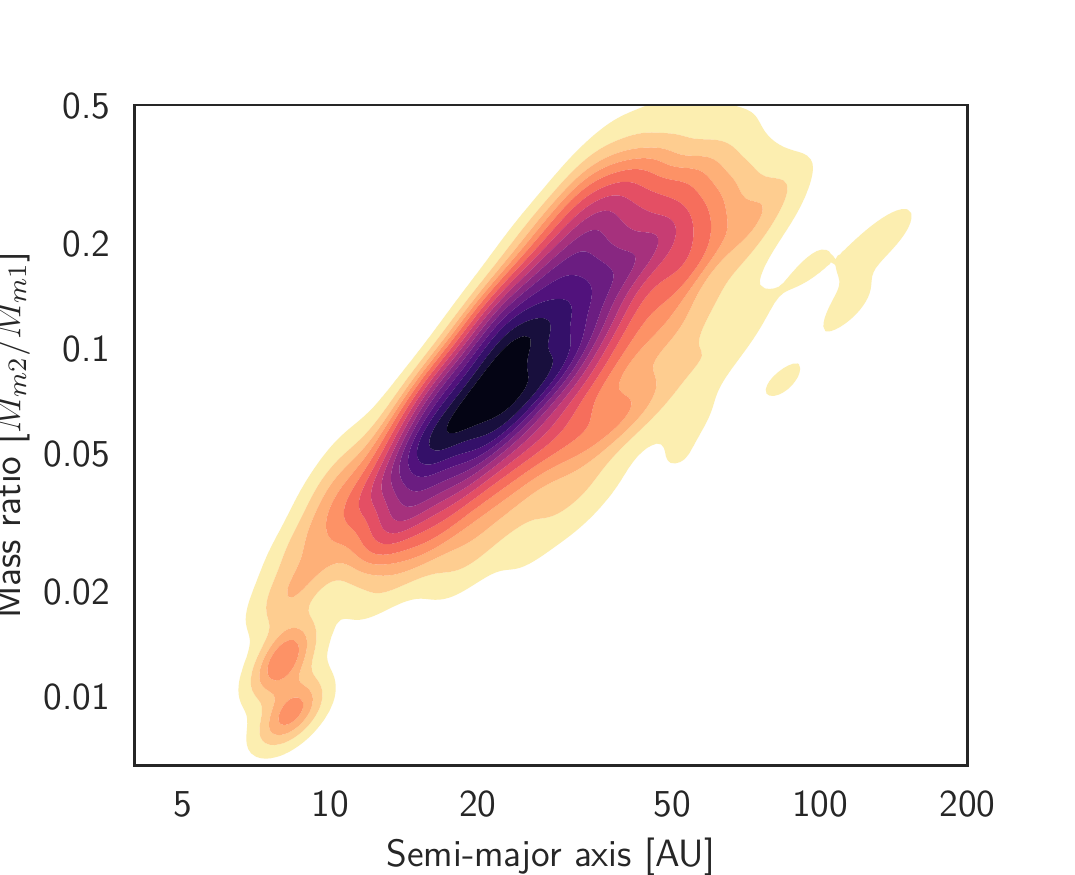}
    \caption{Kernel density function showing the distribution of semi-major axis \& mass ratio for long-period companions compatible with all available observable data.}
    \label{fig:compkde}
\end{figure}

The remaining samples (shown in Figure \ref{fig:compkde}) therefore represent those which are consistent with the RV trend, (lack of) astrometric acceleration, imaging constraints, the lack of extra lines in the combined spectrum, and the assumed stability of the internal multiplanet system.
The companion is therefore likely a brown dwarf or low-mass M dwarf close to the hydrogen burning limit with a mass of $ 0.08^{+0.12}_{-0.05} M_\odot$ and a semi-major axis of $26.0^{+19.0}_{-11.0}$\,AU.

Our analysis of the orbit of \Tstar{}\,B hints that the outer companion (inclination $89\pm6^{\circ}$) is closely aligned with the transiting planets (average inclination of $89.77\pm0.12$).
\refcom{This tightly constrained inclination is due to two factors. Firstly the astrometric shift between Hipparcos and Gaia appears to be linear rather than the two-dimensional shift expected from a face-on orbit. Secondary the model tries to maximise the RV trend - which is substantial - while minimising the secondary mass - which is limited by the strict upper mass limits from the non-detection in high-constrast imaging and the astrometric amplitude.}
This is consistent with observations of small planet transiting systems which are far more likely to host aligned binary companions than the stellar average \citep{Christian2022}.

This approach also allows us to assess the detectability of the companion. In the Ks band, the magnitude difference to the primary star is expected to be only $ 6.8^{+1.2}_{-2.2} $, however we estimate it is currently at a separation of only $ 0.143 \pm 0.05 $arcsec, which \refcom{may make imaging the companion challenging.}

This analysis also rests on the assumption that the observed RV variation is exclusively from a single external companion. 
If, for example, a giant planet and low-mass star both exist on long orbits in this system, then this could also produce the observed RV curvature.
Hence, monitoring of \Tstar{} over a series of years with high precision spectrographs is needed to confirm the hypothesis presented here.

\subsection{Interior Composition} 
Given we have both mass and radius values for \Tstar{}\,b and c, we are able to constrain their mean densities. This allows us to model the internal structure of the planets. We use the method described in \citet{Leleu2021}, which is based on \citet{Dorn2017}. We will only outline the most important aspects of the Bayesian inference model here, namely the observational input parameters, the priors and the main assumptions that are part of the forward model used to calculate the likelihood of the sampled internal structure parameters.

We assume planets that are spherically symmetric with four fully distinct layers: An inner iron core, a silicate mantle, a water layer and a pure H/He atmosphere. Our forward model uses equations of state from \citet{Hakim2018}, \citet{Sotin2007}, \citet{Haldemann2020} and \citet{Lopez2014} to model these layers. Moreover, the current version of our model makes two important assumptions: First, we assume a fixed temperature and pressure at the atmosphere-water boundary and model the gas layer separately from the rest of the planet, which means we neglect any influence of the gas layer on the solid part. Second, we follow \citet{Thiabaud2015} and assume that the Si/Mg/Fe ratios of the planets match the ones of the star exactly.
Furthermore, we model both planets in the system simultaneously.
In a future version of our model, we plan implementing more recent results from \citet{Adibekyan2021}, which shows that while the composition of the star and its planets correlate, they do not necessarily share identical composition.

The model takes as input parameters various planetary and stellar observables, more specifically the transit depths, relative masses and periods of the planets and the mass, radius, age, effective temperature, metallicity and Si and Mg abundances of the star. We assume a prior that is uniform in log for the gas mass fraction. The prior we assume for the layer mass fractions of the iron core, silicate mantle and the water layer (with respect to the solid planet) is uniform, with the added conditions that they need to add up to 1 and the water mass fraction has a maximum value of 0.5 \citep{Thiabaud2014}. 
We stress that the results of our model depend to a certain extent on the chosen priors and repeating the calculation with very different priors might lead to different posterior distributions for the internal structure parameters.

\begin{table}
    \caption{Posterior distributions of the internal structure parameters of \Tstar{}\,b and c. The values correspond to the median and the 5 and 95 percentile of the distributions.}
    \label{tab:int_structure}
    \centering
    \begin{tabular}{lrr}
    \hline\hline                 
    Internal structure parameter & \Tstar{}\,b & \Tstar{}\,c \\
    \hline
    M\textsubscript{core}/M\textsubscript{total} & {\small $0.13^{+0.13}_{-0.12}$} & {\small $0.13^{+0.13}_{-0.11}$} \\
    \noalign{\smallskip}
    M\textsubscript{water}/M\textsubscript{total} & {\small $0.23^{+0.24}_{-0.21}$} & {\small $0.23^{+0.24}_{-0.21}$} \\
    \noalign{\smallskip}
    log M\textsubscript{gas} [M\textsubscript{$\oplus$}] & {\small $-0.15^{+0.19}_{-0.23}$} & {\small $-0.59^{+0.27}_{-0.47}$} \\
    \noalign{\smallskip}
    Fe\textsubscript{core} & {\small $0.90^{+0.09}_{-0.08}$} & {\small $0.90^{+0.09}_{-0.08}$} \\
    \noalign{\smallskip}
    Si\textsubscript{mantle} & {\small $0.40^{+0.08}_{-0.05}$} & {\small $0.40^{+0.08}_{-0.05}$} \\
    \noalign{\smallskip}
    Mg\textsubscript{mantle} & {\small $0.45^{+0.11}_{-0.11}$} & {\small $0.45^{+0.11}_{-0.11}$} \\
    \noalign{\smallskip}
    \hline
    \end{tabular}
\end{table}

The results of our model for both planets are summarised in Table\,\ref{tab:int_structure}. The full corner plots can be found in Appendix\,\ref{sec:appendix_int_struc}. For both \Tstar{}\,b and c, the posterior of the water mass fraction is almost completely unconstrained. Conversely, the posteriors of the gas masses are reasonably well constrained, considering the rather high uncertainty on the mass of planet\,c. For planet\,b, the posterior distribution of the gas mass has a median of $0.71^{+0.40}_{-0.29}$\,M$_{\oplus}$ (error bars are the 5 and 95 percentile of the posterior), corresponding to a thickness of $1.76^{+0.26}_{-0.27}$\,R$_{\oplus}$ of the gas layer. For planet\,c, the median of the gas mass is $0.26^{+0.22}_{-0.17}$\,M$_{\oplus}$, with a thickness of $1.47^{+0.40}_{-0.32}$\,R$_{\oplus}$ of the gas layer.

\subsection{Planet c}
\Tstar{}\,c was re-detected at a 52.5 d period with extremely high confidence in our \cheops{} data. 
We find a transit depth of 1.2ppt which, given the magnitudes of all known nearby stars, cannot be due to a blend.
As discussed in Section \ref{sect:companion}, all available data suggests the bound stellar (or brown dwarf) companion in the system is low-mass and extremely unlikely to be the source of any signal.
The nearby background stellar contaminant discussed in sect \ref{sect:gaia} is also not bright enough to cause the observed transits of planet c.

We clearly see the reflex motion of planet b in our radial velocities, confirming it as a true planet orbiting \Tstar{}.
Thanks to their low mutual inclinations and stability, multi-planet systems have extremely low false positive rates \citep{Lissauer2012}.
We also find, thanks to precise stellar parameters and high-SNR transit observations, that the transit duration matches extremely closely to that from our derived stellar density, implying both that planet c transits our proposed target star, and that its orbit is likely close to circular.
Given all of these considerations, we are therefore confident in calling both \Tstar{}\,b \& c bona fide planets. 

Despite a mass measurement for \Tstar{}\,b we only find a marginal detection of the reflex motion of \Tstar{}\,c in the RVs.
Indeed, our combined model finds a semi-amplitude of only \TKone{},\ms{} and a $3\sigma$ mass limit of $<18$\,\mearth{}.
With a far longer period this could simply be due to the reduction in semi-amplitude with orbital period ($K_p\propto P^{-1/3}$).
However, it appears more likely (\refcom{$\sim66\%$} according to our combined model) that the mass of \Tstar{}\,c is lower than that of \Tstar{}\,b.
\refcom{We also found eccentricities for the two planets consistent with zero (at $<2\sigma$).}

Interestingly, we found \Tstar{}\,c to be orbiting just beyond a 5/2 integer ratio (2.530). 
Although this is a third order resonance, there is still the possibility that TTVs may be present. 
However, when modelling the system using floating individual transit times for each of the 4 transits of \Tstar{}\,b and 3 transits of \Tstar{}\,c and found no obvious TTVs despite timing constrained to $<5$min.

\subsection{Similar systems}
Three other bright ($G<10$) host stars host a system of multiple Neptune-like transiting planets ($3<R_p<5\,R_\oplus$).
TOI-2076 hosts planets at 3.5 \& 3.2$R_\oplus$ on orbits of 21 \& 35d respectively \citep{Osborn2022}.
HD\,28109 hosts planets with 4.2 \& 3.25 $R_\oplus$ on orbits of 56 \& 84\,d \citep{Dransfield2022}.
HD\,191939 hosts three such planets with radii of 3.4, 3.1 and 3.0$R_\oplus$ on orbits of 9, 29 \& 38d \citep{Badenas-Agusti2020}.  
In all of these cases, \Tstar{} included, it appears that these planet's ability to maintain thick atmospheres is likely helped by their longer orbital periods and therefore low insolation.
This is partly because, at longer periods and lower insolations, the typical upper radius limit is not sculpted by the "hot Neptune desert" - a regime where high evaporation causes atmospheric loss.

These planets, all found by \tess{}, are bright enough to allow precise radial velocities, and therefore we have masses for the planets and constraints on the outer companions.
In the majority of these cases, the inner Neptune is the most massive despite not frequently holding the largest radius.
Radial velocity measurements also revealed that HD\,139139 also hosts a high-mass outer companion on a long-period orbit \citep{Lubin2022,Orell-Miquel2022}, much like \Tstar{}.
Whether such companions could have helped produce their similar architectures is an open question.

\begin{figure}
\includegraphics[width=\columnwidth]{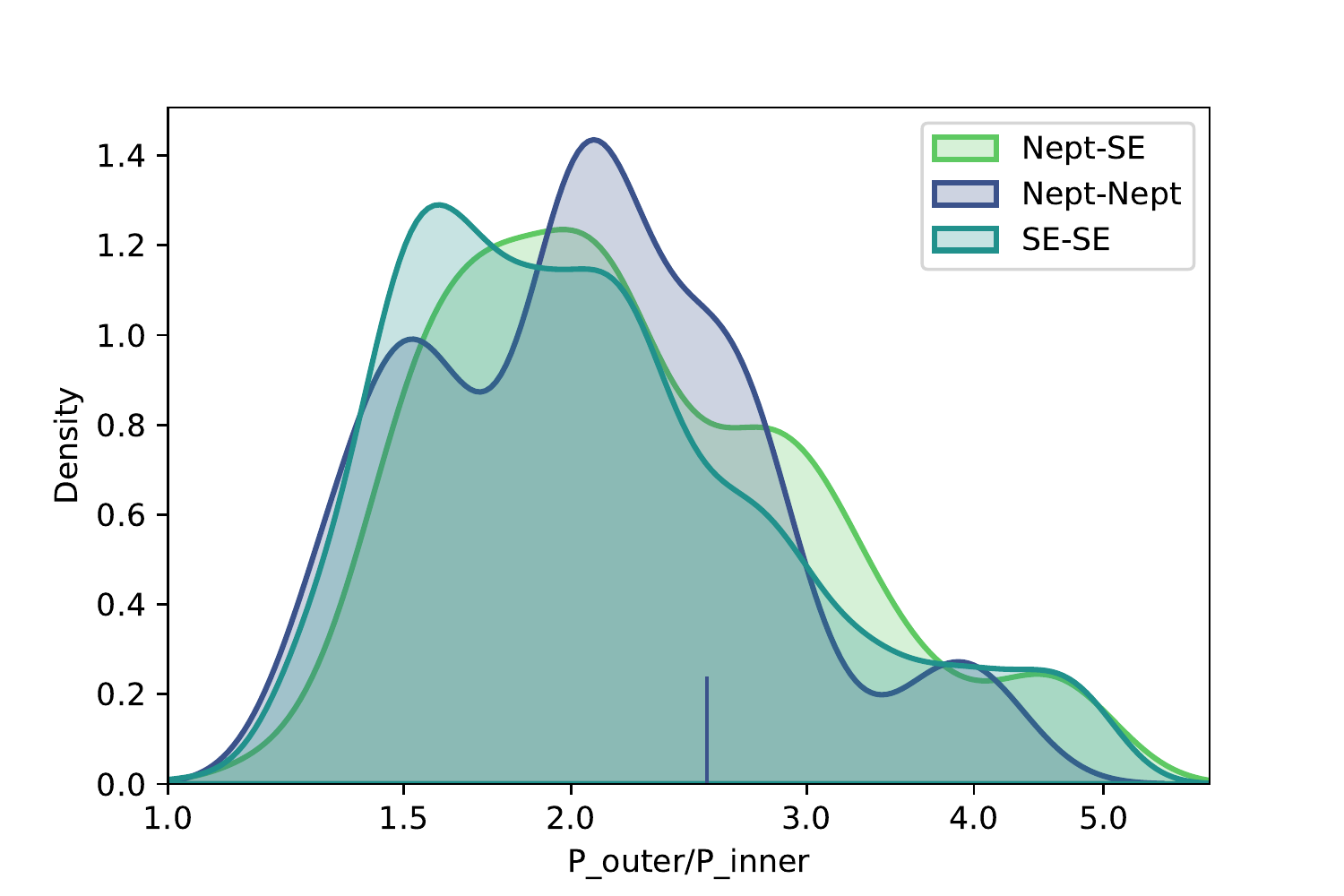}
\caption{Period ratios for neighbouring planets in multi systems split into three groups - Neptune pairs ($3<R_p<5R_\oplus$), super-earth pairs ($R_p<3R_\oplus$), and dissimilar pairs spanning the two. Exoplanets come from the NASA exoplanet archive \citep{Akeson2013}. Ratio populations are displayed using a kernel density estimator \citep[KDE,][]{sheather1991reliable,Waskom2021}. The dash at the bottom shows the position of \Tstar{}}\label{fig:per_ratios}
\end{figure}

To put \Tstar{} into context, we also studied the period-ratios of neighbouring pairs in transiting multi-planet systems. 
\Tstar{} is unusual in that it hosts two planets larger than the typical radius for multiplanet systems - both are closer to Neptune in radius than to typical super Earths ($\sim1.5R_\oplus$) or even mini-Neptunes ($\sim2.2R_\oplus$).
We find that, while all small planets typically have period ratios peaking at $\sim2$, Neptune-Neptune pairs (defined here as planets with $3<R_p<5R_\oplus$) have significantly larger average period ratios than both pairs of small planets ($<3R_\oplus$) and dissimilar pairs (Figure \ref{fig:per_ratios}).
This effect is even stronger when divided by mass rather than radius, however the fact that small planets are dominated by TTV masses and therefore period ratios close to e.g. 1.5 may bias this dataset.
Hence, their larger-than-average planetary sizes may explain why the \Tstar{} planetary system did not end up in (or did not persist as) a closely-orbiting resonant chain, unlike many multi systems of smaller planets.

\subsection{Characterisation Potential}
With its bright IR magnitude (K=7.8), the \Tstar{} system is amenable to atmospheric follow-up.
We calculate Transmission Spectroscopy Metric \citep[TSM,][]{Kempton2018} values of \TTSMzero{} for \Tstar{}\,b using our derived mass and radius, and \TTSMoneshort{} for \Tstar{}\,c.

\begin{figure}
  \includegraphics[width=\columnwidth,trim={20 20 20 20}]{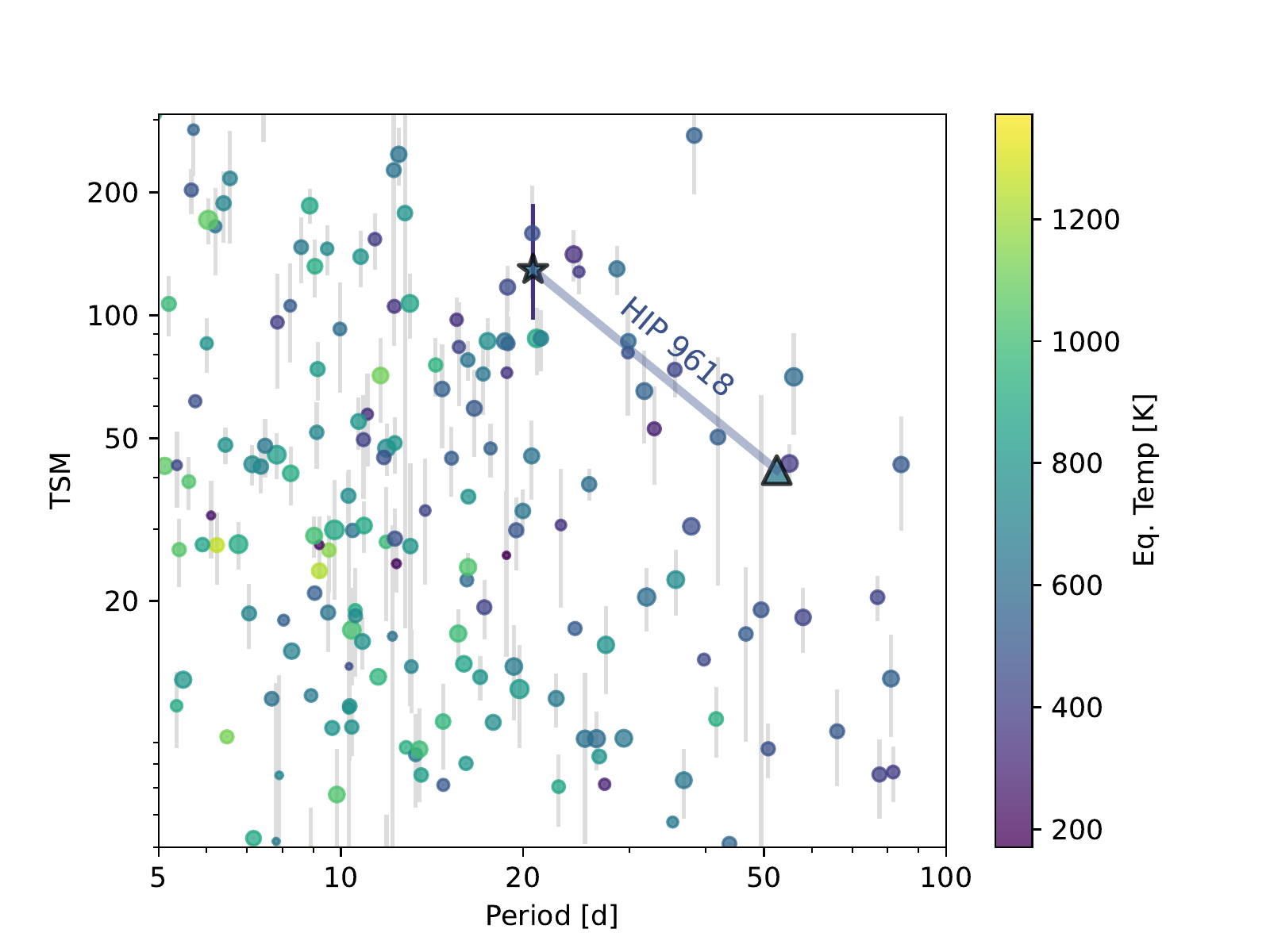}
  \caption{Transmission Spectroscopy Metrics (TSMs) \refcom{and uncertainties} for small ($R_p<5R_\oplus$) exoplanets as a function of orbital period and equilibrium temperature. Exoplanetary data is taken from the NASA Exoplanet Archive \citep{Akeson2013}. \refcom{\Tstar{}\,b is shown as a star \& the 3-sigma lower limit is shown for \Tstar{}\,c as a triangle}.}
  \label{fig:tsms}
\end{figure}

\begin{figure}
  \includegraphics[width=\columnwidth]{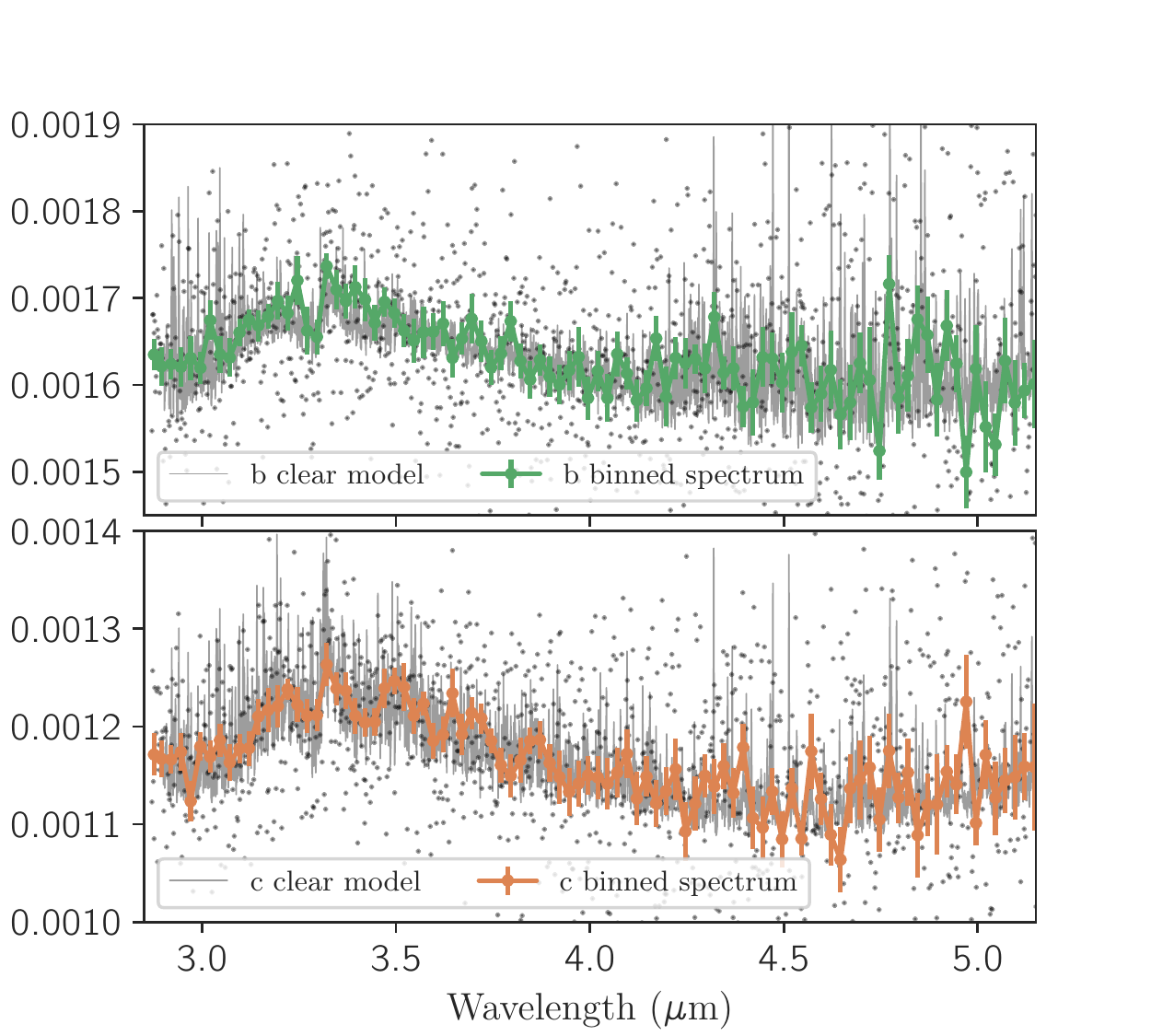}
  \caption{Simulated JWST spectra of \Tstar{}\,b \refcom{(upper panel)} \& c \refcom{(lower panel)} using \texttt{PANDEXO} for \refcom{a clear atmospheric model}. Small dark points are individual JWST simulated high-resolution data, while coloured tickmarks represent the spectra sampled in 25nm bins. Lines represent the input models from \citet{Fortney2010}.}
  \label{fig:jwst_spec}
\end{figure}

Nevertheless, these two planets are some of the most amenable sub-Neptunes on long orbits around solar-like stars (see Figure \ref{fig:tsms}).
To test the observability of spectral features with JWST we used the \texttt{PANDEXO} package \citep{Batalha2017} to produce simulated spectra of the two planets from a single transit.
We chose to test the NIRSpec/BOTS/G395M which is optimal for this target both because saturation is avoided at the redder modes and because the information content is highest in this mode for sub-Neptunes \citep{Guzman-Mesa2020}.
We used our derived planetary parameters as inputs, using the 2-sigma upper limit mass for \Tstar{}\,c.
As model spectra we used cloudy and clear equilibrium chemistry models from \citet{Fortney2010} at 750 \& 500K for \Tstar{}\,b \& \Tstar{}\,c, respectively. Simulated outputs can be seen in Figure \ref{fig:jwst_spec}, where we find spectral features including the CH$_4$ feature at 3.3$\mu$m clearly visible in both planets in both \refcom{cloudless and moderately cloudy models (not shown)}.

\section{Conclusions}
\tess{} has proved exceptional for finding transiting planets around bright stars which are amenable to follow-up including high-resolution spectroscopy and transmission spectroscopy.
\Tstar{}\,b, found by \tess{} to have a period of \Tperiodzero{}\,d and a radius of \Trplzero{}\,\rearth{} is an excellent example of this.
The parent star, an old ($9\pm4$ Gyr) sunlike (\teff$=5611\pm31$) star, has bright visual and infrared magnitudes (G=$9.03$, K=$7.56$) that helped enable precise follow-up RVs which found it to have a mass of \TMpzero{}\,$M_\oplus$.
The precise radius and mass enable internal structure modelling which suggest the planet to have a $6.8\pm1.4\%$ Hydrogen-Helium gaseous envelope and potentially a water-rich core.
With a TSM of $103\pm18$, it is one of the few highly-rated warm ($T_{\rm eq}<750\,\mathrm{K}$ small planets orbiting a solar-like star. Simulations using \texttt{PANDEXO} suggest spectral features could be readily detectable in its atmosphere with a single \JWST{} transit observation.

Unlike \Tstar{}\,b, \tess{} was unable to adequately detect \Tstar{}\,c at the true period.
This is because, for planets on long periods not at the ecliptic poles, \tess{}'s short observation windows (27 \& 54d in the case of \Tstar{}) do not enable it to observe consecutive transits.
In the case of \Tstar{}\,c, \tess{} saw two transits spaced by 680\,d - a ''Duotransit'' candidate with a radius of \Trplone{}\rearth{}.
We modelled the two transits using \texttt{MonoTools} and used \cheops{} to search the highest-probability period aliases, successfully recovering a transit on the fourth observation and finding a period of \Tperiodoneshort{}\,d.
Our RVs do not find a reliable mass for \Tstar{}\,c, but we can place a $3\sigma$ upper limit of \refcom{\TMponeshort{}$M_\oplus$}.
This suggests that planet c is less massive than b.
Establishing a true mass may be possible with future RV measurements, which would in turn be key to enable accurate atmospheric characterisation.
Even using a conservative upper mass limit, it is likely that \Tstar{}\,c becomes one of the few characterisable planets orbiting a Sun-like star with an equilibrium temperature below $500~K$.

Due to its bright magnitude, archival RV measurements exist for \Tstar{} taken with the \sophie{} spectrograph starting 8 years before the detection of the candidate planets. This displays a clear $\sim200$~\ms{} long-term trend consistent with a massive outer companion.
Combined with the 20-year baseline of \textit{Hipparcos} and \gaia{} astrometric measurements, and high-resolution imaging follow-up observations, we are able to constrain the mass and orbit of this outer companion, finding that a companion near the hydrogen burning limit orbiting at 15-50 au best fits the observations.
This companion can be better constrained with long-term RV monitoring, data from the \gaia{} extended mission, and potentially even with targeted direct imaging in the IR \citep[e.g.][]{Bonavita2022}. 

This detection makes \Tstar{} one of only five bright (K<8) transiting multi-planet systems which hosts a planet with P>50d, opening the door for atmospheric characterisation of a regime of warm ($T_{\rm eq}<750$K) sub-Neptunes.

\section*{Acknowledgements}
\scriptsize{
We thank the TESS-Keck Survey team for their open coordination on this target, especially Joey Murphy, Howard Isaacson, Erik Petigura, Andrew Howard, \& Natalie Batalha.

This paper includes data collected by the \tess{} mission. Funding for the \tess{} mission is provided by the NASA Explorer Program. We acknowledge the use of public TOI Release data from pipelines at the \tess{} Science Office and at the \tess{} Science Processing Operations Center. Resources supporting this work were provided by the NASA High-End Computing (HEC) Program through the NASA Advanced Supercomputing (NAS) Division at Ames Research Center for the production of the SPOC data products. This research has made use of the Exoplanet Follow-up Observation Program (ExoFOP; DOI: 10.26134/ExoFOP5) website, which is operated by the California Institute of Technology, under contract with the National Aeronautics and Space Administration under the Exoplanet Exploration Program.
This work makes use of observations from the LCOGT network. Part of the LCOGT telescope time was granted by NOIRLab through the Mid-Scale Innovations Program (MSIP). MSIP is funded by NSF.
Some of the observations in the paper made use of the High-Resolution Imaging instrument ‘Alopeke obtained under Gemini LLP Proposal Number: GN/S-2021A-LP-105. ‘Alopeke was funded by the NASA Exoplanet Exploration Program and built at the NASA Ames Research Center by Steve B. Howell, Nic Scott, Elliott P. Horch, and Emmett Quigley. Alopeke was mounted on the Gemini North (and/or South) telescope of the international Gemini Observatory, a program of NSF’s OIR Lab, which is managed by the Association of Universities for Research in Astronomy (AURA) under a cooperative agreement with the National Science Foundation. on behalf of the Gemini partnership: the National Science Foundation (United States), National Research Council (Canada), Agencia Nacional de Investigación y Desarrollo (Chile), Ministerio de Ciencia, Tecnología e Innovación (Argentina), Ministério da Ciência, Tecnologia, Inovações e Comunicações (Brazil), and Korea Astronomy and Space Science Institute (Republic of Korea).
This work has made use of data from the European Space Agency (ESA) mission {\it Gaia} (\url{https://www.cosmos.esa.int/gaia}), processed by the {\it Gaia} Data Processing and Analysis Consortium (DPAC, \url{https://www.cosmos.esa.int/web/gaia/dpac/consortium}). Funding for the DPAC has been provided by national institutions, in particular the institutions participating in the {\it Gaia} Multilateral Agreement.
The Digitized Sky Surveys were produced at the Space Telescope Science Institute under U.S. Government grant NAG W-2166. The images of these surveys are based on photographic data obtained using the Oschin Schmidt Telescope on Palomar Mountain and the UK Schmidt Telescope. The plates were processed into the present compressed digital form with the permission of these institutions. The National Geographic Society - Palomar Observatory Sky Atlas (POSS-I) was made by the California Institute of Technology with grants from the National Geographic Society. The Second Palomar Observatory Sky Survey (POSS-II) was made by the California Institute of Technology with funds from the National Science Foundation, the National Geographic Society, the Sloan Foundation, the Samuel Oschin Foundation, and the Eastman Kodak Corporation. 
Based on observations made with the Italian Telescopio Nazionale Galileo (TNG) operated on the island of La Palma by the Fundaci\'on Galileo Galilei of the INAF (Istituto Nazionale di Astrofisica) at the Spanish Observatorio del Roque de los Muchachos of the Instituto de Astrofisica de Canarias under programs ITP19\_1 and CAT21B\_39.
Based on observations collected at the Centro Astron\'omico Hispano en Andaluc\'ia (CAHA) at Calar Alto, operated jointly by Junta de Andaluc\'ia and Consejo Superior de Investigaciones Cient\'ificas (IAA-CSIC).
This work has been carried out within the framework of the NCCR PlanetS supported by the Swiss National Science Foundation under grants 51NF40\_182901 and 51NF40\_205606.\\ 
G.N. thanks for the research funding from the Ministry of Education and Science programme the "Excellence Initiative - Research University" conducted at the Centre of Excellence in Astrophysics and Astrochemistry of the Nicolaus Copernicus University in Toru\'n, Poland.\\ 
\\ 
ABr was supported by the SNSA.\\ 
ACC acknowledges support from STFC consolidated grant numbers ST/R000824/1 and ST/V000861/1, and UKSA grant number ST/R003203/1.\\ 
JAEg acknowledges support from the Swiss National Science Foundation (SNSF) under grant 200020\_192038.\\ 
DG gratefully acknowledges financial support from the CRT foundation under Grant No. 2018.2323 ``Gaseousor rocky? Unveiling the nature of small worlds''.\\ 
ML acknowledges support of the Swiss National Science Foundation under grant number PCEFP2\_194576.\\ 
ACC and TW acknowledge support from STFC consolidated grant numbers ST/R000824/1 and ST/V000861/1, and UKSA grant number ST/R003203/1.\\ 
YAl acknowledges support from the Swiss National Science Foundation (SNSF) under grant 200020\_192038.\\ 
We acknowledge support from the Spanish Ministry of Science and Innovation and the European Regional Development Fund through grants ESP2016-80435-C2-1-R, ESP2016-80435-C2-2-R, PGC2018-098153-B-C33, PGC2018-098153-B-C31, ESP2017-87676-C5-1-R, MDM-2017-0737 Unidad de Excelencia Maria de Maeztu-Centro de Astrobiologí­a (INTA-CSIC), as well as the support of the Generalitat de Catalunya/CERCA programme. The MOC activities have been supported by the ESA contract No. 4000124370.\\ 
A.A.B. \& M.V.G. acknowledge the support of Ministry of Science and Higher Education of the Russian Federation under the grant 075-15-2020-780 (N13.1902.21.0039)\\ 
We warmly thank the OHP staff for their support on the observations. X.B., I.B. and T.F. received funding from the French Programme National de Physique Stellaire (PNPS) and the Programme National de Planétologie (PNP) of CNRS (INSU).\\ 
XB, SC, DG, MF and JL acknowledge their role as ESA-appointed CHEOPS science team members.\\ 
This project was supported by the CNES\\ 
The Belgian participation to CHEOPS has been supported by the Belgian Federal Science Policy Office (BELSPO) in the framework of the PRODEX Program, and by the University of Liège through an ARC grant for Concerted Research Actions financed by the Wallonia-Brussels Federation; L.D. is an F.R.S.-FNRS Postdoctoral Researcher.\\ 
B.-O. D. acknowledges support from the Swiss State Secretariat for Education, Research and Innovation (SERI) under contract number MB22.00046.\\ 
This project has received funding from the European Research Council (ERC) under the European Union’s Horizon 2020 research and innovation programme (project {\sc Four Aces}; grant agreement No 724427). It has also been carried out in the frame of the National Centre for Competence in Research PlanetS supported by the Swiss National Science Foundation (SNSF). DE acknowledges financial support from the Swiss National Science Foundation for project 200021\_200726.\\ 
MF and CMP gratefully acknowledge the support of the Swedish National Space Agency (DNR 65/19, 174/18).\\ 
M.G. is an F.R.S.-FNRS Senior Research Associate.\\ 
MNG is the ESA CHEOPS Project Scientist and Mission Representative, and as such also responsible for the Guest Observers (GO) Programme. MNG does not relay proprietary information between the GO and Guaranteed Time Observation (GTO) Programmes, and does not decide on the definition and target selection of the GTO Programme.\\ 
SH gratefully acknowledges CNES funding through the grant 837319.\\ 
KGI is the ESA CHEOPS Project Scientist and is responsible for the ESA CHEOPS Guest Observers Programme. She does not participate in, or contribute to, the definition of the Guaranteed Time Programme of the CHEOPS mission through which observations described in this paper have been taken, nor to any aspect of target selection for the programme.\\ 
J.~K. gratefully acknowledges the support of the Swedish National Space Agency (SNSA; DNR 2020-00104) and of the Swedish Research Council (VR: Etableringsbidrag 2017-04945)\\ 
PM acknowledges support from STFC research grant number ST/M001040/1.\\ 
This work was granted access to the HPC resources of MesoPSL financed by the Region Ile de France and the project Equip@Meso (reference ANR-10-EQPX-29-01) of the programme Investissements d'Avenir supervised by the Agence Nationale pour la Recherche\\ 
EM acknowledges funding from the French National Research Agency (ANR) under contract number ANR\-18\-CE31\-0019 (SPlaSH) and funding from Fundação de Amparo à Pesquisa do Estado de Minas Gerais (FAPEMIG) under the project number APQ\-02493\-22.\\ 
\\ 
LBo, GBr, VNa, IPa, GPi, RRa, GSc, VSi, and TZi acknowledge support from CHEOPS ASI-INAF agreement n. 2019-29-HH.0.\\ 
\\ 
This work was also partially supported by a grant from the Simons Foundation (PI Queloz, grant number 327127).\\ 
\\ 
IRI acknowledges support from the Spanish Ministry of Science and Innovation and the European Regional Development Fund through grant PGC2018-098153-B- C33, as well as the support of the Generalitat de Catalunya/CERCA programme.\\ 
This work was supported by FCT - Fundação para a Ciência e a Tecnologia through national funds and by FEDER through COMPETE2020 - Programa Operacional Competitividade e Internacionalizacão by these grants: UID/FIS/04434/2019, UIDB/04434/2020, UIDP/04434/2020, PTDC/FIS-AST/32113/2017 \& POCI-01-0145-FEDER- 032113, PTDC/FIS-AST/28953/2017 \& POCI-01-0145-FEDER-028953, PTDC/FIS-AST/28987/2017 \& POCI-01-0145-FEDER-028987, O.D.S.D. is supported in the form of work contract (DL 57/2016/CP1364/CT0004) funded by national funds through FCT.\\ 
NCSa acknowledges funding by the European Union (ERC, FIERCE, 101052347). Views and opinions expressed are however those of the author(s) only and do not necessarily reflect those of the European Union or the European Research Council. Neither the European Union nor the granting authority can be held responsible for them.\\ 
\\ 
S.G.S. acknowledge support from FCT through FCT contract nr. CEECIND/00826/2018 and POPH/FSE (EC)\\ 
GyMSz acknowledges the support of the Hungarian National Research, Development and Innovation Office (NKFIH) grant K-125015, a a PRODEX Experiment Agreement No. 4000137122, the Lend\"ulet LP2018-7/2021 grant of the Hungarian Academy of Science and the support of the city of Szombathely.\\ 
V.V.G. is an F.R.S-FNRS Research Associate.\\ 
NAW acknowledges UKSA grant ST/R004838/1\\ 
}
\section*{Data Availability}

The TESS data presented here is publicly accessible at MAST at \url{http://mast.stsci.edu}. All CHEOPS data presented here will be uploaded to CDS on publication of this paper \url{https://cds.u-strasbg.fr/}. Radial velocities are included in the tables found in the appendix. Ground-based photometry and imaging is publicly available through ExoFOP at \url{https://exofop.ipac.caltech.edu/tess/}



\bibliographystyle{mnras}
\bibliography{_PAPERrefs} 



\appendix

\section*{Affiliations}
\label{sec:affiliations}

%
%
\textsuperscript{\hypertarget{affil_1}{1}} Physikalisches Institut, University of Bern, Gesellsschaftstrasse 6, 3012 Bern, Switzerland\\
\textsuperscript{\hypertarget{affil_2}{2}} Department of Physics and Kavli Institute for Astrophysics and Space Research, Massachusetts Institute of Technology, Cambridge, MA 02139, USA\\
\textsuperscript{\hypertarget{affil_3}{3}} Institute of Astronomy, Faculty of Physics, Astronomy and Informatics, Nicolaus Copernicus University, Grudzi\c{a}dzka 5, 87-100 Toru\'n, Poland \\
\textsuperscript{\hypertarget{affil_4}{4}} Instituto de Astrof\'isica de Canarias, C/ v\'ia L\'actea, s/n, E-38205 La Laguna, Tenerife, Spain\\
\textsuperscript{\hypertarget{affil_5}{5}} Departamento de Astrof\'isica, Universidad de La Laguna, Av. Astrof\'isico Francisco S\'anchez, s/n, E-38206 La Laguna, Tenerife, Spain\\
\textsuperscript{\hypertarget{affil_6}{6}} Institut d'astrophysique de Paris, UMR7095 CNRS, Universit\'e Pierre \& Marie Curie, 98bis boulevard Arago, 75014 Paris, France\\
\textsuperscript{\hypertarget{affil_7}{7}} Centro de Astrobiolog\'ia (CAB,CSIC-INTA), Dep. de Astrof\'isica, ESAC campus, 28692, Villanueva de la Ca\~nada, Madrid, Spain\\
\textsuperscript{\hypertarget{affil_8}{8}} Observatoire Astronomique de l'Université de Genève, Chemin Pegasi 51, 1290 Versoix, Switzerland\\
\textsuperscript{\hypertarget{affil_9}{9}} Department of Astronomy, Stockholm University, AlbaNova University Center, 10691 Stockholm, Sweden\\
\textsuperscript{\hypertarget{affil_10}{10}} LAM, Aix Marseille Univ, CNRS, CNES, Marseille, France\\
\textsuperscript{\hypertarget{affil_11}{11}} Instituto de Astrof\'isica e Ci\^encias do Espa\c{c}o, Universidade do Porto, CAUP, Rua das Estrelas, 4150-762 Porto, Portugal\\
\textsuperscript{\hypertarget{affil_12}{12}} Center for Astrophysics \textbar Harvard \& Smithsonian, 60 Garden Street, Cambridge, MA 02138, USA\\
\textsuperscript{\hypertarget{affil_13}{13}} INAF, Osservatorio Astronomico di Padova, Vicolo dell'Osservatorio 5, 35122 Padova, Italy\\
\textsuperscript{\hypertarget{affil_14}{14}} NASA Exoplanet Science Institute - Caltech/IPAC, Pasadena, CA 91125 USA\\
\textsuperscript{\hypertarget{affil_15}{15}} Centre for Exoplanet Science, SUPA School of Physics and Astronomy, University of St Andrews, North Haugh, St Andrews KY16 9SS, UK\\
\textsuperscript{\hypertarget{affil_16}{16}} Dipartimento di Fisica, Universita degli Studi di Torino, via Pietro Giuria 1, I-10125, Torino, Italy\\
\textsuperscript{\hypertarget{affil_17}{17}} Astrophysics Group, Cavendish Laboratory, University of Cambridge, J.J. Thomson Avenue, Cambridge CB3 0HE, UK\\
\textsuperscript{\hypertarget{affil_18}{18}} Max-Planck-Institut für Astronomie, Königstuhl 17, 69117 Heidelberg, Germany\\
\textsuperscript{\hypertarget{affil_19}{19}} Department of Physics, Engineering and Astronomy, Stephen F. Austin State University, 1936 North St, Nacogdoches, TX 75962, USA\\
\textsuperscript{\hypertarget{affil_20}{20}} Institut de Ciencies de l'Espai (ICE, CSIC), Campus UAB, Can Magrans s/n, 08193 Bellaterra, Spain\\
\textsuperscript{\hypertarget{affil_21}{21}} Institut d'Estudis Espacials de Catalunya (IEEC), 08034 Barcelona, Spain\\
\textsuperscript{\hypertarget{affil_22}{22}} Observatoire de Haute-Provence, CNRS, Universit\'e d'Aix-Marseille, 04870 Saint-Michel-l'Observatoire, France\\
\textsuperscript{\hypertarget{affil_23}{23}} European Space Agency (ESA), European Space Research and Technology Centre (ESTEC), Keplerlaan 1, 2201 AZ Noordwijk, The Netherlands\\
\textsuperscript{\hypertarget{affil_24}{24}} Depto. de Astrofisica, Centro de Astrobiologia (CSIC-INTA), ESAC campus, 28692 Villanueva de la Cañada (Madrid), Spain\\
\textsuperscript{\hypertarget{affil_25}{25}} Space Research Institute, Austrian Academy of Sciences, Schmiedlstrasse 6, A-8042 Graz, Austria\\
\textsuperscript{\hypertarget{affil_26}{26}} Sternberg Astronomical Institute Lomonosov Moscow State University, Universitetskii prospekt, 13, Moscow, 119992, Russia\\
\textsuperscript{\hypertarget{affil_27}{27}} Center for Space and Habitability, University of Bern, Gesellschaftsstrasse 6, 3012, Bern, Switzerland\\
\textsuperscript{\hypertarget{affil_28}{28}} Max Planck Institute for Extraterrestrial Physics\\
\textsuperscript{\hypertarget{affil_29}{29}} Université Grenoble Alpes, CNRS, IPAG, 38000 Grenoble, France\\
\textsuperscript{\hypertarget{affil_30}{30}} 1DTU Space, National Space Institute, Technical University of Denmark, Elektrovej 328, DK-2800 Kgs. Lyngby, Denmark\\
\textsuperscript{\hypertarget{affil_31}{31}} Admatis, 5. Kandó Kálmán Street, 3534 Miskolc, Hungary\\
\textsuperscript{\hypertarget{affil_32}{32}} Instituto de Astrofisica e Ciencias do Espaco, Universidade do Porto, CAUP, Rua das Estrelas, 4150-762 Porto, Portugal\\
\textsuperscript{\hypertarget{affil_33}{33}} Departamento de Fisica e Astronomia, Faculdade de Ciencias, Universidade do Porto, Rua do Campo Alegre, 4169-007 Porto, Portugal\\
\textsuperscript{\hypertarget{affil_34}{34}} Institute of Planetary Research, German Aerospace Center (DLR), Rutherfordstrasse 2, 12489 Berlin, Germany\\
\textsuperscript{\hypertarget{affil_35}{35}} Université de Paris, Institut de physique du globe de Paris, CNRS, F-75005 Paris, France\\
\textsuperscript{\hypertarget{affil_36}{36}} Centre for Mathematical Sciences, Lund University, Box 118, 22100 Lund, Sweden\\
\textsuperscript{\hypertarget{affil_37}{37}} Universit\'e Grenoble Alpes, CNRS, IPAG, 38000 Grenoble, France\\
\textsuperscript{\hypertarget{affil_38}{38}} Astrobiology Research Unit, Université de Liège, Allée du 6 Août 19C, B-4000 Liège, Belgium\\
\textsuperscript{\hypertarget{affil_39}{39}} Space sciences, Technologies and Astrophysics Research (STAR) Institute, Universit\'e de Li\`ege, 19C All\'ee du 6 Ao\^ut, B-4000 Li\`ege, Belgium\\
\textsuperscript{\hypertarget{affil_40}{40}} Center for Space and Habitability, Gesellsschaftstrasse 6, 3012 Bern, Switzerland\\
\textsuperscript{\hypertarget{affil_41}{41}} Lunar and Planetary Laboratory, The University of Arizona, Tucson, AZ 85721, USA\\
\textsuperscript{\hypertarget{affil_42}{42}} Leiden Observatory, University of Leiden, PO Box 9513, 2300 RA Leiden, The Netherlands\\
\textsuperscript{\hypertarget{affil_43}{43}} Department of Space, Earth and Environment, Chalmers University of Technology, Onsala Space Observatory, 43992 Onsala, Sweden\\
\textsuperscript{\hypertarget{affil_44}{44}} Department of Astronomy and Astrophysics, University of California, Santa Cruz, CA 95064, USA\\
\textsuperscript{\hypertarget{affil_45}{45}} Department of Astrophysics, University of Vienna, Tuerkenschanzstrasse 17, 1180 Vienna, Austria\\
\textsuperscript{\hypertarget{affil_46}{46}} NASA Ames Research Center, Moffett Field, CA 94035, USA\\
\textsuperscript{\hypertarget{affil_47}{47}} LESIA, Observatoire de Paris, Université PSL, Sorbonne Université, Université Paris Cité, CNRS, 5 place Jules Janssen, 92195 Meudon, France\\
\textsuperscript{\hypertarget{affil_48}{48}} Konkoly Observatory, Research Centre for Astronomy and Earth Sciences, 1121 Budapest, Konkoly Thege Miklós út 15-17, Hungary\\
\textsuperscript{\hypertarget{affil_49}{49}} ELTE E\"otv\"os Lor\'and University, Institute of Physics, P\'azm\'any P\'eter s\'et\'any 1/A, 1117 Budapest, Hungary\\
\textsuperscript{\hypertarget{affil_50}{50}} Department of Space, Earth and Environment, Astronomy and Plasma Physics, Chalmers University of Technology, 412 96 Gothenburg, Sweden.\\
\textsuperscript{\hypertarget{affil_51}{51}} Lund Observatory, Division of Astrophysics, Department of Physics, Lund University, Box 43, 22100 Lund, Sweden\\
\textsuperscript{\hypertarget{affil_52}{52}} Astrophysics Group, Keele University, Staffordshire, ST5 5BG, United Kingdom\\
\textsuperscript{\hypertarget{affil_53}{53}} IMCCE, UMR8028 CNRS, Observatoire de Paris, PSL Univ., Sorbonne Univ., 77 av. Denfert-Rochereau, 75014 Paris, France\\
\textsuperscript{\hypertarget{affil_54}{54}} Institut d'astrophysique de Paris, UMR7095 CNRS, Université Pierre \& Marie Curie, 98bis blvd. Arago, 75014 Paris, France\\
\textsuperscript{\hypertarget{affil_55}{55}} Department of Astronomy \& Astrophysics, University of Chicago, Chicago, IL 60637, USA.\\
\textsuperscript{\hypertarget{affil_56}{56}} Laborat\'{o}rio Nacional de Astrof\'{i}sica, Rua Estados Unidos 154, 37504-364, Itajub\'{a} - MG, Brazil\\
\textsuperscript{\hypertarget{affil_57}{57}} Space Telescope Science Institute, 3700 San Martin Drive, Baltimore, MD, 21218, USA\\
\textsuperscript{\hypertarget{affil_58}{58}} Google, Cambridge, MA, USA\\
\textsuperscript{\hypertarget{affil_59}{59}} Universit\'e de Toulouse, CNRS, IRAP, 14 avenue Belin, 31400 Toulouse, France\\
\textsuperscript{\hypertarget{affil_60}{60}} Department of Physics \& Astronomy, Swarthmore College, Swarthmore PA 19081, USA\\
\textsuperscript{\hypertarget{affil_61}{61}} INAF, Osservatorio Astrofisico di Catania, Via S. Sofia 78, 95123 Catania, Italy\\
\textsuperscript{\hypertarget{affil_62}{62}} Dipartimento di Fisica e Astronomia "Galileo Galilei", Universita degli Studi di Padova, Vicolo dell'Osservatorio 3, 35122 Padova, Italy\\
\textsuperscript{\hypertarget{affil_63}{63}} Department of Physics, University of Warwick, Gibbet Hill Road, Coventry CV4 7AL, United Kingdom\\
\textsuperscript{\hypertarget{affil_64}{64}} Centre for Origin and Prevalence of Life, ETH Zurich, Wolfgang-Pauli-Strasse 27, 8093 Zurich, Switzerland\\
\textsuperscript{\hypertarget{affil_65}{65}} Cavendish Laboratory, JJ Thomson Avenue, Cambridge CB3 0HE, UK\\
\textsuperscript{\hypertarget{affil_66}{66}} Center for Astronomy and Astrophysics, Technical University Berlin, Hardenberstrasse 36, 10623 Berlin, Germany\\
\textsuperscript{\hypertarget{affil_67}{67}} Institut für Geologische Wissenschaften, Freie UniversitÃ¤t Berlin, 12249 Berlin, Germany\\
\textsuperscript{\hypertarget{affil_68}{68}} ELTE E\"otv\"os Lor\'and University, Gothard Astrophysical Observatory, 9700 Szombathely, Szent Imre h. u. 112, Hungary\\
\textsuperscript{\hypertarget{affil_69}{69}} MTA-ELTE Exoplanet Research Group, 9700 Szombathely, Szent Imre h. u. 112, Hungary\\
\textsuperscript{\hypertarget{affil_70}{70}} German Aerospace Center (DLR), Institute of Optical Sensor Systems, Rutherfordstraße 2, 12489 Berlin\\
\textsuperscript{\hypertarget{affil_71}{71}} Institute of Astronomy, University of Cambridge, Madingley Road, Cambridge, CB3 0HA, United Kingdom\\
\textsuperscript{\hypertarget{affil_72}{72}} Department of Astrophysical Sciences, Princeton University, 4 Ivy Lane, Princeton, NJ 08544, USA\\

\section{Detailed description of high-resolution imaging observations}

\subsubsection{Keck \& Palomar}\label{sect:Keck_palomar} 
The Keck Observatory observations were made with the NIRC2 instrument on Keck-II behind the natural guide star AO system \citep{wizinowich2000} on 2020-May-28 UT in a standard 3-point dither pattern that is used with NIRC2 to avoid the left lower quadrant of the detector which shows excess noise. The dither pattern step size was $3\arcsec$ and was repeated twice, with each dither offset from the previous dither by $0.5\arcsec$. 
NIRC2 was used in the narrow-angle mode with a full field of view of $\sim10\arcsec$ and a pixel scale of approximately $0.0099442\arcsec$ per pixel. The Keck observations were made in the narrow-band filters $Br-\gamma$ filter $(\lambda_o = 2.1686; \Delta\lambda = 0.0326~\mu$m) with an integration time of 2 seconds for a total of 18 seconds on target.

The Palomar Observatory observations of \Tstar{} were made with the PHARO instrument \citep{hayward2001} behind the natural guide star AO system P3K \citep{dekany2013} on 2020~Dec~04 in a standard 5-point \texttt{quincunx} dither pattern with steps of 5\arcsec\ in the narrow-band $Br-\gamma$ filter $(\lambda_o = 2.1686; \Delta\lambda = 0.0326~\mu$m).  Each dither position was observed three times, offset in position from each other by 0.5\arcsec\ for a total of 15 frames; with an integration time of 1.4 seconds per frame, respectively for total on-source times of 21 seconds. PHARO has a pixel scale of $0.025\arcsec$ per pixel for a total field of view of $\sim25\arcsec$.

Palomar \& Keck data were processed and analyzed with a custom set of IDL tools. The science frames were flat-fielded and sky-subtracted.  The flat fields were generated from a median average of dark subtracted flats taken on-sky.  The flats were normalized such that the median value of the flats is unity.  The sky frames were generated from the median average of the dithered science frames; each science image was then sky-subtracted and flat-fielded. Reduced science frames were combined into a single combined image using a intra-pixel interpolation that conserves flux, shifts the individual dithered frames by the appropriate fractional pixels, and median-coadds the frames.  The final resolutions of the combined dithers were determined from the full-width half-maximum of the point spread functions: 0.12\arcsec\ and 0.078\arcsec\ for the Palomar and Keck observations, respectively.  

\begin{figure}
\centering
   \begin{subfigure}[h]{0.45\textwidth}
    \includegraphics[width=\textwidth, trim=50 12 0 0]{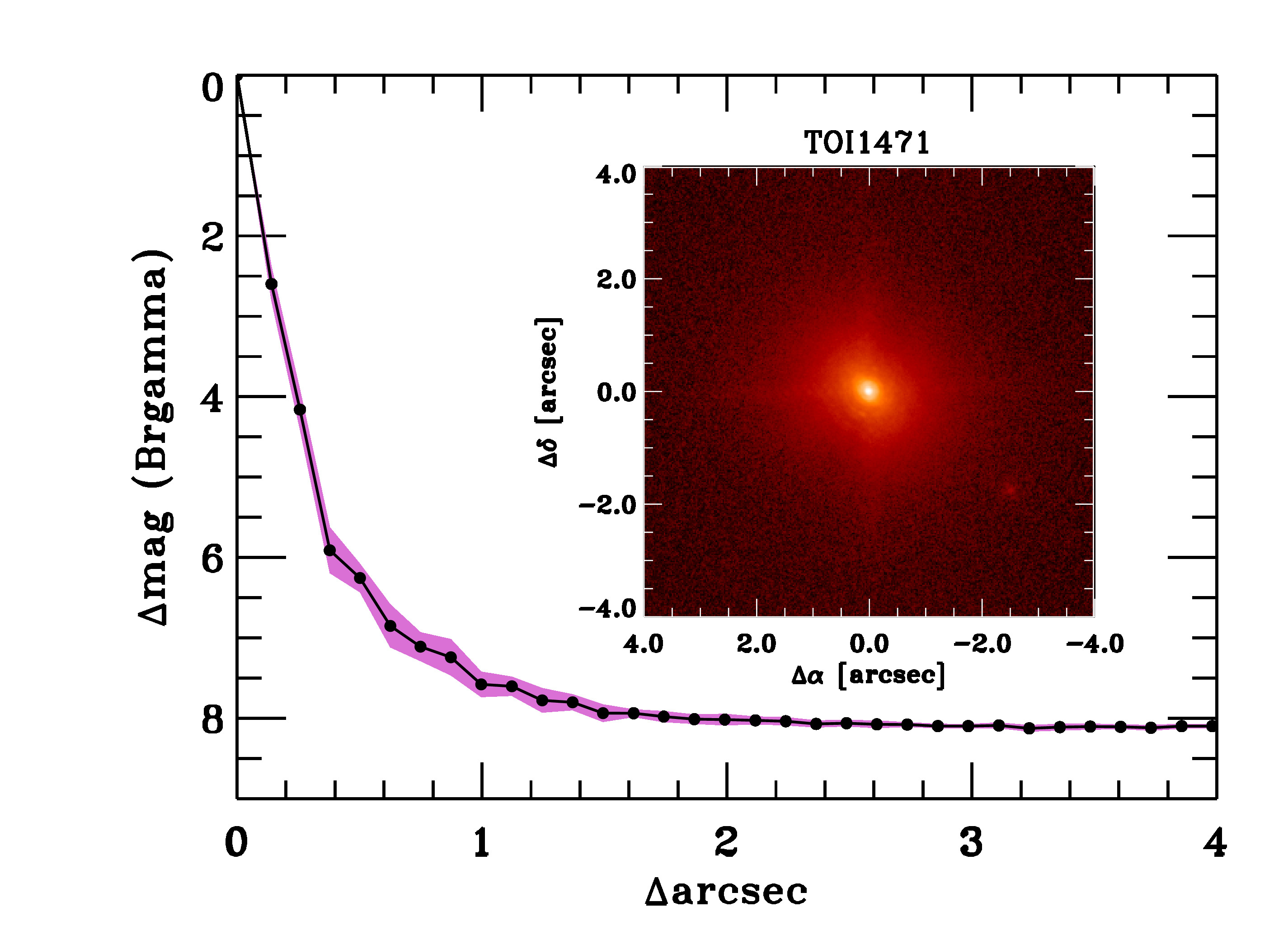}
   \caption{Palomar NIR AO imaging and sensitivity curve for \Tstar{} (or TOI-1471) taken in the Br$\gamma$ filter. The image reaches a contrast of $\sim 7$ magnitudes fainter than the host star within 0.\arcsec5. {\it Inset:} Image of the central portion of the data, centered on the star and showing the faint background star 3\arcsec\ to the southwest.}
   \label{fig:imagingpal} 
\end{subfigure}
\begin{subfigure}[h]{0.45\textwidth}
    \includegraphics[width=0.9\textwidth]{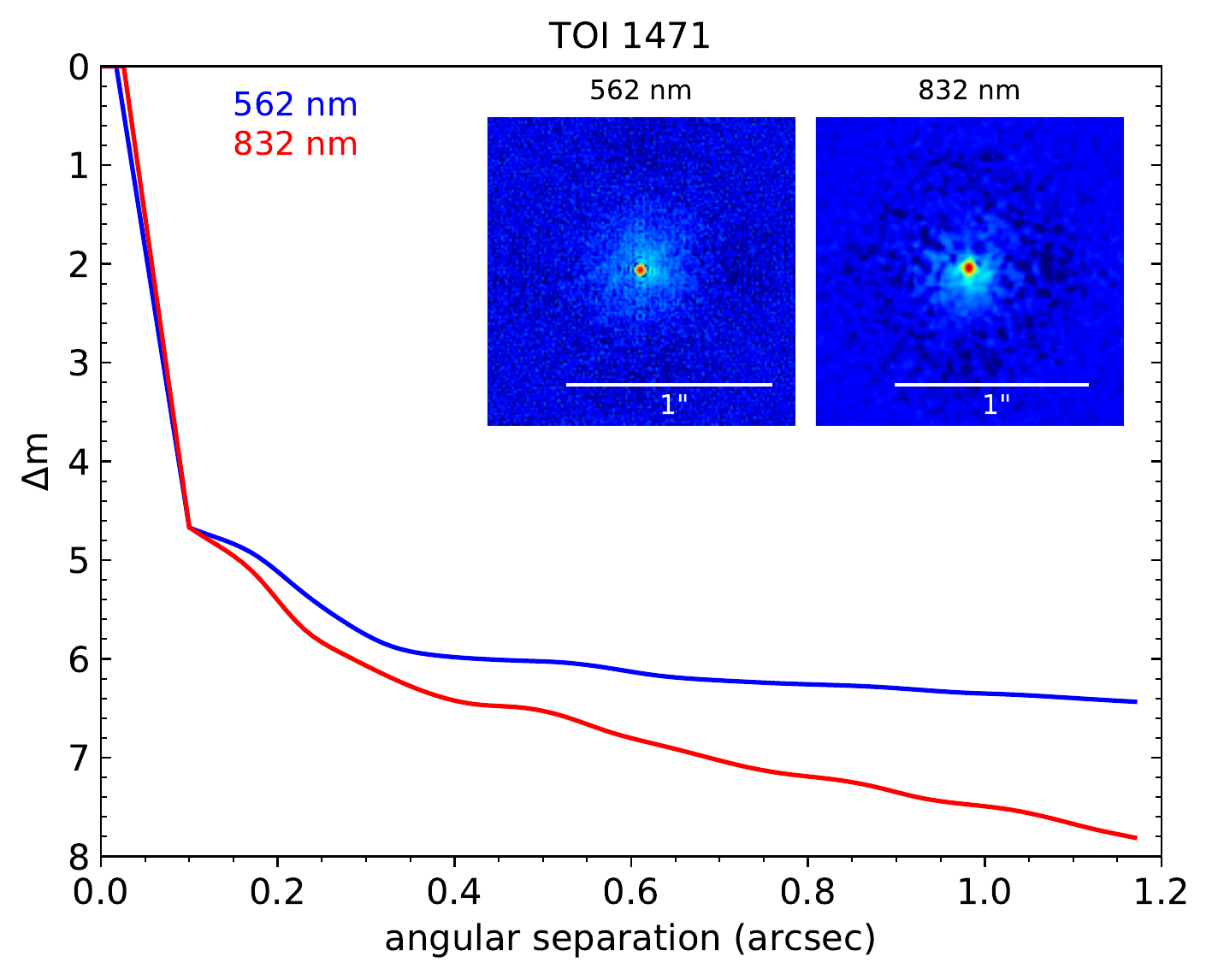}
   \caption{We show the $5\sigma$ speckle imaging contrast curves in both filters as a function of the angular separation from the diffraction limit out to 1.2 arcsec. The inset shows the reconstructed 832 nm image with a 1 arcsec scale bar. The star, \Tstar{} or TOI1471, was found to have no close companions to within the angular and contrast levels achieved.}
   \label{fig:imaginggem}
\end{subfigure}
\begin{subfigure}[h]{0.45\textwidth}
    \includegraphics[width=0.9\textwidth]{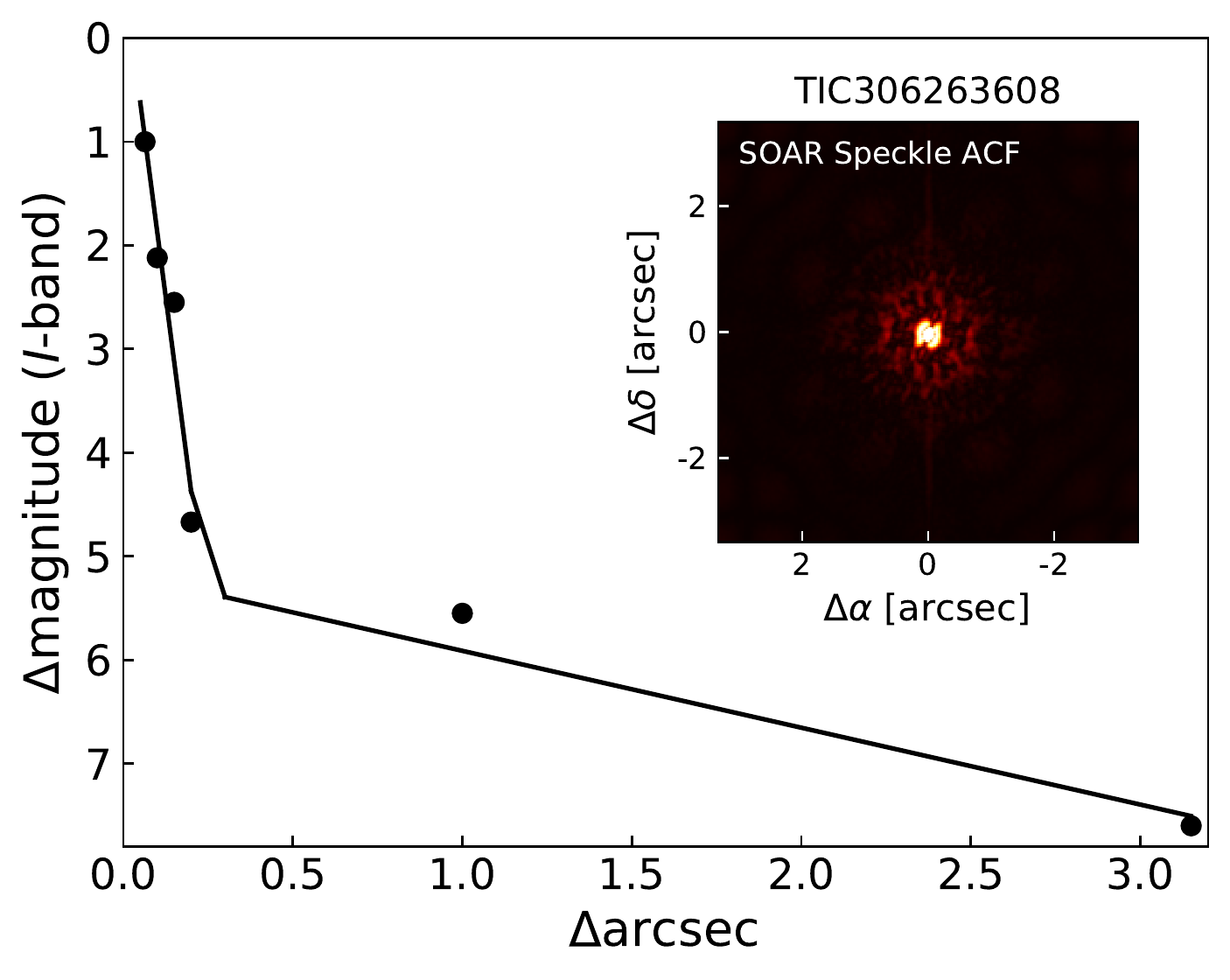}
   \caption{Similar to above, the speckle imaging contrast curve from the SOAR telescope taken in visible I-band filter. The inset shows the speckle auto-correlation function, showing no nearby star within the sensitivity of the observation.}
   \label{fig:imagingsoar}
\end{subfigure}
\caption{High-resolution imaging results.}
\end{figure}

The sensitivities of the final combined AO image were determined by injecting simulated sources azimuthally around the primary target every $20^\circ $ at separations of integer multiples of the central source's FWHM \citep{furlan2017}. The brightness of each injected source was scaled until standard aperture photometry detected it with $5\sigma $ significance. The resulting brightness of the injected sources relative to \Tstar{} set the contrast limits at that injection location. The final $5\sigma $ limit at each separation was determined from the average of all of the determined limits at that separation and the uncertainty on the limit was set by the rms dispersion of the azimuthal slices at a given radial distance.  The Keck data have better sensitivity closer-in ($\delta \rm{mag} = 2.5$ mag at 0.1\arcsec), but the Palomar data are deeper sensitivity at wider separations ($\delta \rm{mag} = 8$ mag at 2\arcsec); the final sensitivity curves for the Palomar is shown in Figure~\ref{fig:imagingpal}.
    
\subsubsection{Gemini}\label{sect:Gemini} 
TOI-1471 was observed on 2020 December 04 UT and 2021 October 17 UT using the ‘Alopeke speckle instrument on the Gemini North 8-m telescope \citep{SH,HF}.  ‘Alopeke provides simultaneous speckle imaging in two bands (562nm and 832 nm) with output data products including a reconstructed image with robust contrast limits on companion detections. While both observations had consistent results that TOI-1471 is a single star to within the angular and contrast levels achieved, the October 2021 observation had better seeing which led to deeper contrast levels. Twelve sets of $1000 \times 0.06$ second images were obtained and processed in our standard reduction pipeline \citep{Howell2011}. Figure \ref{fig:imaginggem} shows our final contrast curves and the 832 nm reconstructed speckle image. We find that TOI-1471 is a single star with no companion brighter than 5-8 magnitudes below that of the target star from the 8-m telescope diffraction limit (20 mas) out to 1.2”. At the distance of TOI-1471 (d=67.5 pc) these angular limits correspond to spatial limits of 1.35 to 81 AU.

\subsubsection{SOAR}\label{sect:SOAR} 

We searched for nearby stellar companions to TOI-1471 with speckle imaging on the 4.1-m Southern Astrophysical Research (SOAR) telescope \citep{tokovinin2018} on 31 October 2020 UT, observing in Cousins I-band, a similar visible bandpass as TESS. This observation was sensitive to a 7.6-magnitude fainter star at an angular distance of 1 arcsec from the target. More details of the observations within the SOAR TESS survey are available in \citet{ziegler2020}. The 5$\sigma$ detection sensitivity and speckle auto-correlation functions from the observations are shown in Figure \ref{fig:imagingsoar} . No nearby stars were detected within 3\arcsec of TOI-1471 in the SOAR observations.

\subsubsection{SAI-2.5m}\label{sect:SAI} 

We also observed TOI-1471 on 3 December 2020 UT with the SPeckle Polarimeter \citet[SPP][]{Safonov2017} on the 2.5 m telescope at the Caucasian Observatory of Sternberg Astronomical Institute (SAI) of Lomonosov Moscow State University in the spectral band centered on 625~nm with FWHM of 50~nm. SPP uses an Electron Multiplying CCD Andor iXon 897 as a detector, and we used the atmospheric dispersion compensation. The detector has a pixel scale of 20.6 mas/pixel, the angular resolution is 89 mas, and the field of view is $5\arcsec\times5\arcsec$ centered on the star. The power spectrum was estimated from 4000 frames with 30 ms exposures. We did not detect any stellar companions brighter than $\Delta$mag = 4.3 and 6.2 at $0.2\arcsec$ and $0.5\arcsec$, respectively.

\section{RV observations}
\begin{table*}
\begin{center}
\caption{Radial velocities and spectral activity indicators measured from TNG/HARPS-N spectra with the DRS.
\label{tab:harpsn_rvs}}
\begin{tabular}{rrrrrrrrrrr}
\hline
\hline
\multicolumn{1}{r}{BJD$_\mathrm{TBD}$} &
\multicolumn{1}{r}{RV} &
\multicolumn{1}{r}{$\sigma_\mathrm{RV}$} &
\multicolumn{1}{r}{BIS} &
\multicolumn{1}{r}{$\sigma_\mathrm{BIS}$} &
\multicolumn{1}{r}{CCF\_FWHM} &
\multicolumn{1}{r}{CCF\_CTR} &
\multicolumn{1}{r}{$\mathrm{\log{R^{`}_{HK}}}$} &
\multicolumn{1}{r}{$\mathrm{\sigma_{\log{R^{`}_{HK}}}}$} &
\multicolumn{1}{r}{SNR@550nm} &
\multicolumn{1}{r}{$\mathrm{T_{exp}}$}\\ 
\multicolumn{1}{r}{-2457000} &
\multicolumn{1}{r}{($\mathrm{m\,s^{-1}}$)} &
\multicolumn{1}{r}{($\mathrm{m\,s^{-1}}$)} &
\multicolumn{1}{r}{($\mathrm{m\,s^{-1}}$)} &
\multicolumn{1}{r}{($\mathrm{m\,s^{-1}}$)} &
\multicolumn{1}{r}{($\mathrm{km\,s^{-1}}$)} &
\multicolumn{1}{r}{(\%)} &
\multicolumn{1}{r}{---} &
\multicolumn{1}{r}{---} &
\multicolumn{1}{r}{(@550nm)} &
\multicolumn{1}{r}{(s)}\\ 
\hline
     1869.31908 &      -16047.975 &           1.897 &         -27.281 &           2.682 &           6.780 &          47.865 &         -4.9633 &          0.0205 &            47.9 &           281.1\\
     1869.33575 &      -16033.064 &           0.784 &         -25.601 &           1.109 &           6.779 &          47.942 &         -4.9847 &          0.0060 &           105.3 &          1500.0\\
     1895.39675 &      -16042.441 &           1.373 &         -27.726 &           1.941 &           6.781 &          48.022 &         -5.0056 &          0.0160 &            63.7 &          1500.0\\
     1896.38630 &      -16040.879 &           1.049 &         -24.883 &           1.484 &           6.787 &          47.982 &         -4.9759 &          0.0097 &            81.1 &          1500.0\\
     1897.38259 &      -16041.553 &           0.938 &         -22.277 &           1.326 &           6.777 &          47.980 &         -4.9577 &          0.0078 &            90.5 &          1500.0\\
     1898.38722 &      -16040.521 &           0.783 &         -25.694 &           1.108 &           6.783 &          47.944 &         -4.9640 &          0.0059 &           107.3 &          1500.0\\
     1905.36412 &      -16043.817 &           1.567 &         -28.000 &           2.216 &           6.778 &          47.951 &         -4.9860 &          0.0191 &            57.5 &          1723.9\\
     2548.36087 &      -16080.145 &           1.945 &         -33.469 &           2.750 &           6.768 &          48.124 &         -5.0687 &          0.0392 &            51.1 &          2400.0\\
     2574.43126 &      -16091.191 &           4.260 &         -35.382 &           6.024 &           6.777 &          47.795 &         -4.8758 &          0.0726 &            28.4 &           680.8\\
     2574.45186 &      -16081.583 &           1.483 &         -32.075 &           2.098 &           6.768 &          47.980 &         -4.9197 &          0.0205 &            66.6 &          1800.0\\
     2594.45356 &      -16081.331 &           1.521 &         -29.413 &           2.151 &           6.769 &          47.996 &         -4.9653 &          0.0215 &            63.3 &          1200.0\\
     2609.38503 &      -16085.546 &           2.553 &         -29.236 &           3.610 &           6.776 &          48.181 &         -5.0135 &          0.0504 &            40.9 &           600.0\\
     2610.39501 &      -16082.079 &           1.138 &         -33.068 &           1.609 &           6.772 &          48.079 &         -4.9536 &          0.0119 &            81.8 &           600.0\\
\hline
\end{tabular}
\end{center}
\end{table*}

\begin{table}
    \centering
    \caption{\sophie{} Radial Velocities}
    \begin{tabular}{c c c}
        \hline
        BJD [JD-2400000] & RV [\ms{}] & $\sigma_{\rm RV}$ [\ms{}] \\
        \hline
          $ 55853.53644 $ & $ -15914.0 $ & $ 2.5 $ \\
          $ 55883.46881 $ & $ -15930.8 $ & $ 2.7 $ \\
          $ 55916.41221 $ & $ -15914.6 $ & $ 3.2 $ \\
          $ 58840.4519 $ & $ -16074.0 $ & $ 3.7 $ \\
          $ 58841.37571 $ & $ -16071.4 $ & $ 3.9 $ \\
          $ 58857.29082 $ & $ -16070.8 $ & $ 3.8 $ \\
          $ 58858.37755 $ & $ -16077.1 $ & $ 3.8 $ \\
          $ 58887.31391 $ & $ -16044.8 $ & $ 3.8 $ \\
          $ 59057.60564 $ & $ -16085.0 $ & $ 1.6 $ \\
          $ 59058.60044 $ & $ -16084.1 $ & $ 3.9 $ \\
          $ 59060.60405 $ & $ -16100.6 $ & $ 3.8 $ \\
          $ 59063.63157 $ & $ -16074.4 $ & $ 3.8 $ \\
          $ 59082.57844 $ & $ -16072.7 $ & $ 3.8 $ \\
          $ 59112.60148 $ & $ -16094.3 $ & $ 3.9 $ \\
          $ 59113.64083 $ & $ -16067.6 $ & $ 3.7 $ \\
          $ 59134.53817 $ & $ -16076.2 $ & $ 3.8 $ \\
          $ 59138.52924 $ & $ -16081.4 $ & $ 3.8 $ \\
          $ 59139.49069 $ & $ -16088.6 $ & $ 3.7 $ \\
          $ 59141.49017 $ & $ -16076.4 $ & $ 3.8 $ \\
          $ 59141.49331 $ & $ -16070.1 $ & $ 3.8 $ \\
          $ 59146.51885 $ & $ -16087.6 $ & $ 2.4 $ \\
          $ 59151.48282 $ & $ -16081.6 $ & $ 2.4 $ \\
          $ 59170.48772 $ & $ -16092.2 $ & $ 2.4 $ \\
          $ 59457.6237 $ & $ -16119.7 $ & $ 2.4 $ \\
          $ 59527.52674 $ & $ -16110.2 $ & $ 2.4 $ \\
          $ 59542.47712 $ & $ -16113.9 $ & $ 2.4 $ \\
          $ 59565.33575 $ & $ -16114.2 $ & $ 2.4 $ \\
        \hline    \end{tabular}
    \label{tab:sophie_rvs}
\end{table}

\begin{table}
    \centering
    \caption{\cafe{} Radial Velocities}
    \begin{tabular}{c c c}
        \hline
        BJD & RV [\ms{}] & $\sigma_{\rm RV}$ [\ms{}] \\
        \hline
        $ 2459529.4721648 $ & $ -16184.7 $ & $ 9.9 $ \\
        $ 2459530.4696387 $ & $ -16228.9 $ & $ 8.2 $ \\
        $ 2459549.4579368 $ & $ -16231.9 $ & $ 7.0 $ \\
        $ 2459577.4456190 $ & $ -16212.1 $ & $ 5.7 $ \\
        $ 2459578.4636742 $ & $ -16229.2 $ & $ 6.1 $ \\
        $ 2459617.3591001 $ & $ -16201.4 $ & $ 6.5 $ \\
        $ 2459808.6769329 $ & $ -16200.3 $ & $ 5.4 $ \\
        $ 2459809.6736999 $ & $ -16206.2 $ & $ 5.4 $ \\
        $ 2459810.6673429 $ & $ -16225.4 $ & $ 5.9 $ \\
        \hline    \end{tabular}
    \label{tab:cafe_rvs}
\end{table}
\section{Model Priors \& Posteriors}

\makeatletter
\begin{table*}
\small
\caption{Model parameters, priors, and posteriors for the Combined model.} 
\label{table:appendix_P35_model}      
\centering                    
\begin{tabular}{l c c}        
    \hline\hline                 
    Parameter & Prior & Posterior \\
    \hline                        
    Stellar temperature, \teff{}  [K]  & $\mathcal{N}(\mu=5610,\sigma=30) $ &  $ 5609.0 \pm 33.0 $  \\
Stellar radius, $R_s$ [$R_\oplus$] & $\mathcal{N}(\mu=0.9662,\sigma=0.005) $ &  $ 0.9662 \pm 0.005 $  \\
log stellar surface gravity, \logg{}  [cgs] & $\mathcal{N}(\mu=4.47,\sigma=0.1) $ &  $ 4.45^{+0.016}_{-0.033} $  \\
Log radius ratio, $\log{R_{p,b}/R_s}$ , b & $\mathcal{U}(a=-6.908,b=-2.303)$ &  $ -3.297 \pm 0.01 $  \\
Log radius ratio, $\log{R_{p,c}/R_s}$ , c & $\mathcal{U}(a=-6.908,b=-2.303)$ &  $ -3.451 \pm 0.011 $  \\
Impact parameter, $b_{b}$  & $ \mathcal{{U}}(a=0.0,b=1+R_{p,b}/R_s)^\dagger{}$ &  $ 0.13 \pm 0.11 $  \\
Impact parameter, $b_{c}$  & $ \mathcal{{U}}(a=0.0,b=1+R_{p,c}/R_s)^\dagger{}$ &  $ 0.235^{+0.085}_{-0.062} $  \\
RV semi-amplitude, $K_{b}$ [\ms{}], b & $\mathcal{U}(a=-10,b=10)$ &  $ 2.48 \pm 0.71 $  \\
RV semi-amplitude, $K_{c}$ [\ms{}], c & $\mathcal{U}(a=-10,b=10)$ &  $ 1.39 \pm 0.58 $  \\
Quadratic LD, $u_{{\rm TESS},0} $  & $\mathcal{N}_{\mathcal{U}}(a=0.0,b=1.0,\mu=0.3299,\sigma=0.0500)$ &  $ 0.302 \pm 0.061 $  \\
Quadratic LD, $u_{{\rm TESS},1} $  & $\mathcal{N}_{\mathcal{U}}(a=0.0,b=1.0,\mu=0.2605,\sigma=0.0500)$ &  $ 0.223 \pm 0.086 $  \\
Quadratic LD, $u_{{\rm Cheops},0} $  & $\mathcal{N}_{\mathcal{U}}(a=0.0,b=1.0,\mu=0.4494,\sigma=0.0500)$ &  $ 0.501 \pm 0.064 $  \\
Quadratic LD, $u_{{\rm Cheops},1} $  & $\mathcal{N}_{\mathcal{U}}(a=0.0,b=1.0,\mu=0.2292,\sigma=0.0500)$ &  $ 0.174 \pm 0.082 $  \\
Eccentricity, $e_{b}$  & $ \mathcal{B}{(\alpha=0.867, \beta=3.03)}^{\ddagger}$ &  $ 0.22^{+0.25}_{-0.15} $  \\
Eccentricity, $e_{c}$  & $ \mathcal{B}{(\alpha=0.867, \beta=3.03)}^{\ddagger}$ &  $ 0.23^{+0.25}_{-0.16} $  \\
Argument of periasteron, $\omega_{b}$,  & $\mathcal{{U}}(a=-\pi,b=\pi)$ &  $ 0.0 \pm 3.9 $  \\
Argument of periasteron, $\omega_{c}$,  & $\mathcal{{U}}(a=-\pi,b=\pi)$ &  $ 0.1 \pm 3.9 $  \\
SOPHIE RV mean, $\mu_{\rm SOPHIE}$ & $\mathcal{N}(\mu=-16077.1,\sigma=53.265)$ &  $ -16084.7 \pm 1.9 $  \\
HARPS-N RV mean, $\mu_{\rm HARPS-N}$ & $\mathcal{N}(\mu=-2.8479,\sigma=10.7575)$ &  $ -14.16 \pm 0.36 $  \\
CAFE RV mean, $\mu_{\rm CAFE}$ & $\mathcal{N}(\mu=-16212.1,\sigma=96.2822)$ &  $ -16178.4 \pm 7.1 $  \\
RV quadratic trend, [m.s$^{-1}$.d$^{-2}$] & $\mathcal{N}(\mu=0,\sigma=0.005)$ &  $ -4.99e-06 \pm 6.4e-07 $  \\
RV linear trend, [m.s$^{-1}$.d$^{-1}$] & $\mathcal{N}(\mu=0,\sigma=0.05)$ &  $ -0.0666 \pm 0.0013 $  \\
log RV scatter, $\log{\sigma_{\rm SOPHIE}}$ & $\mathcal{N}(\mu=2.15791,\sigma=2)$ &  $ 2.13 \pm 0.17 $  \\
log RV scatter, $\log{\sigma_{\rm HARPS-N}}$ & $\mathcal{N}(\mu=-1.88719,\sigma=2)$ &  $ -2.2^{+1.5}_{-1.9} $  \\
log RV scatter, $\log{\sigma_{\rm CAFE}}$ & $\mathcal{N}(\mu=2.84818,\sigma=2)$ &  $ 2.98 \pm 0.29 $  \\
Log photometric scatter, $\log{\sigma_{{\rm TESS},s}/(\rm{ppt})}$  & $\mathcal{N}(\mu=0.1435,\sigma=3) $ &  $ 0.1 \pm 3.0 $  \\
I mag contamination from companion, $\Delta I$  & $\mathcal{N}(\mu=7.398,\sigma=1.767) $ &  $ 7.4 \pm 1.7 $  \\
Log photometric scatter, $\log{\sigma_{{\rm Cheops},s}/(\rm{ppt})}$  & $\mathcal{N}(\mu=-7.605,\sigma=3) $ &  $ -7.5 \pm 3.0 $  \\
V mag contamination from companion, $\Delta V$  & $\mathcal{N}(\mu=8.966,\sigma=2.076) $ &  $ 9.0 \pm 2.1 $  \\
Correlation with normalised background flux, $df/d({\cos{(({\rm BG}-\mu_{\rm BG})/\sigma_{\rm BG})}})$   & $\mathcal{N}(\mu=0,\sigma=0.4474) $ &  $ -0.09 \pm 0.1 $  \\
Correlation with normalised sine of roll angle, $df/d({\sin{((\Phi-\mu_{\Phi})/\sigma_{\Phi})}})$   & $\mathcal{N}(\mu=0,\sigma=0.4474) $ &  $ 0.002 \pm 0.048 $  \\
Correlation with normalised cosine of roll angle, $df/d({\cos{((\Phi-\mu_{\Phi})/\sigma_{\Phi})}})$   & $\mathcal{N}(\mu=0,\sigma=0.4474) $ &  $ 0.125 \pm 0.047 $  \\
CHEOPS linear trend for fk=PR110048\_TG017201\_V0200 & $\mathcal{N}(\mu=0,\sigma=0.852144)$ &  $ -0.042 \pm 0.044 $  \\
CHEOPS offset for fk=PR110048\_TG017201\_V0200 & $\mathcal{N}(\mu=0,\sigma=0.313486)$ &  $ 0.01 \pm 0.13 $  \\
CHEOPS linear trend for fk=PR110048\_TG017301\_V0200 & $\mathcal{N}(\mu=0,\sigma=0.852144)$ &  $ 0.009 \pm 0.035 $  \\
CHEOPS offset for fk=PR110048\_TG017301\_V0200 & $\mathcal{N}(\mu=0,\sigma=0.313486)$ &  $ -0.02 \pm 0.13 $  \\
CHEOPS linear trend for fk=PR110048\_TG017401\_V0200 & $\mathcal{N}(\mu=0,\sigma=0.852144)$ &  $ -0.132 \pm 0.038 $  \\
CHEOPS offset for fk=PR110048\_TG017401\_V0200 & $\mathcal{N}(\mu=0,\sigma=0.313486)$ &  $ -0.02 \pm 0.13 $  \\
CHEOPS linear trend for fk=PR110048\_TG017601\_V0200 & $\mathcal{N}(\mu=0,\sigma=0.852144)$ &  $ -0.061 \pm 0.035 $  \\
CHEOPS offset for fk=PR110048\_TG017601\_V0200 & $\mathcal{N}(\mu=0,\sigma=0.313486)$ &  $ -0.02 \pm 0.13 $  \\
BSpline 1 , $F(\Phi)_{{\rm BS},1}$   & $\mathcal{N}(\mu=0.0,\sigma=0.0)$ &  $ 0.153 \pm 0.096 $  \\
BSpline 2 , $F(\Phi)_{{\rm BS},2}$   & $\mathcal{N}(\mu=0.0,\sigma=0.0)$ &  $ -0.004 \pm 0.096 $  \\
BSpline 3 , $F(\Phi)_{{\rm BS},3}$   & $\mathcal{N}(\mu=0.0,\sigma=0.0)$ &  $ 0.014^{+0.095}_{-0.095} $  \\
BSpline 4 , $F(\Phi)_{{\rm BS},4}$   & $\mathcal{N}(\mu=0.0,\sigma=0.0)$ &  $ 0.061 \pm 0.088 $  \\
BSpline 5 , $F(\Phi)_{{\rm BS},5}$   & $\mathcal{N}(\mu=0.0,\sigma=0.0)$ &  $ -0.03 \pm 0.084 $  \\
BSpline 6 , $F(\Phi)_{{\rm BS},6}$   & $\mathcal{N}(\mu=0.0,\sigma=0.0)$ &  $ -0.14 \pm 0.079 $  \\
BSpline 7 , $F(\Phi)_{{\rm BS},7}$   & $\mathcal{N}(\mu=0.0,\sigma=0.0)$ &  $ -0.09^{+0.075}_{-0.074} $  \\
BSpline 8 , $F(\Phi)_{{\rm BS},8}$   & $\mathcal{N}(\mu=0.0,\sigma=0.0)$ &  $ -0.049 \pm 0.071 $  \\
BSpline 9 , $F(\Phi)_{{\rm BS},9}$   & $\mathcal{N}(\mu=0.0,\sigma=0.0)$ &  $ 0.087 \pm 0.071 $  \\
BSpline 10 , $F(\Phi)_{{\rm BS},10}$   & $\mathcal{N}(\mu=0.0,\sigma=0.0)$ &  $ -0.003 \pm 0.072 $  \\
BSpline 11 , $F(\Phi)_{{\rm BS},11}$   & $\mathcal{N}(\mu=0.0,\sigma=0.0)$ &  $ 0.057 \pm 0.074 $  \\
BSpline 12 , $F(\Phi)_{{\rm BS},12}$   & $\mathcal{N}(\mu=0.0,\sigma=0.0)$ &  $ -0.028 \pm 0.077 $  \\
BSpline 13 , $F(\Phi)_{{\rm BS},13}$   & $\mathcal{N}(\mu=0.0,\sigma=0.0)$ &  $ 0.029 \pm 0.08 $  \\
BSpline 14 , $F(\Phi)_{{\rm BS},14}$   & $\mathcal{N}(\mu=0.0,\sigma=0.0)$ &  $ -0.019 \pm 0.089 $  \\
BSpline 15 , $F(\Phi)_{{\rm BS},15}$   & $\mathcal{N}(\mu=0.0,\sigma=0.0)$ &  $ 0.065^{+0.095}_{-0.093} $  \\
BSpline 16 , $F(\Phi)_{{\rm BS},16}$   & $\mathcal{N}(\mu=0.0,\sigma=0.0)$ &  $ -0.12 \pm 0.11 $  \\

    \hline                                   
\end{tabular}
    \begin{tablenotes}
      \footnotesize
      \item $\mathcal{N}$ details a normally distributed prior with mean, $\mu$ and standard deviation, $\sigma$ values. $\mathcal{U}$ details a uniform distribution with lower, $a$, and upper, $b$, limits. $\mathcal{N}_\mathcal{U}$ details a truncated normal distribution with $\mu$,$\sigma$, $a$ \& $b$ values.$\dagger{}$ represents the \citet{espinoza2018efficient} prior and $\ddagger{}$ represents the uniform prior as presented by \citet{2018RNAAS...2..209E} and implemented by \texttt{exoplanet}. ${\ddagger}$ is the prior from \citet{kipping2013}. \cheops{} suffixes refer chronologically to the four unique \cheops{} visits.
    \end{tablenotes}

\end{table*}
\makeatother

\section{Interior Composition}
\label{sec:appendix_int_struc}

\begin{figure}
  \includegraphics[width=\columnwidth]{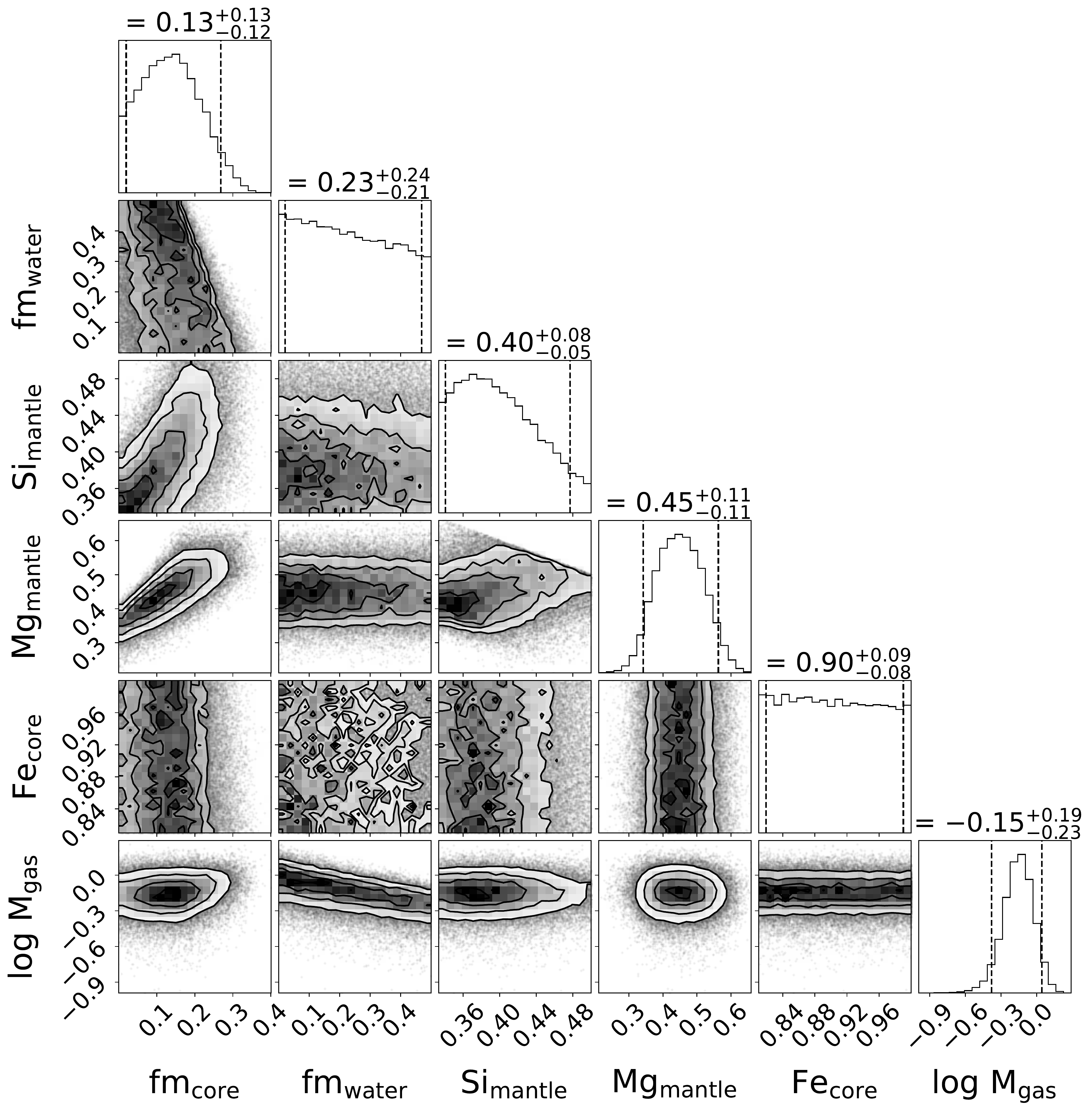}
  \caption{Corner plot showing the posterior distributions of the main internal structure parameters of \Tstar{}\,b: The layer mass fractions of the inner iron core and the water layer (both with respect to the solid planet without the gas layer), the molar fractions of Si and Mg in the mantle and Fe in the iron core and the total gas mass of the planet in Earth masses on a logarithmic scale. The titles of each column correspond to the median and the 5 and 95 percentiles, which are also shown with the dashed lines.}
  \label{fig:int_structure_b}
\end{figure}

\bsp	
\label{lastpage}
\end{document}